\documentclass[9pt,twoside]{pnas-new}
\templatetype{preprintgeneral}
\usepackage{physics}
\usepackage{bm}
\usepackage{microtype}

\title{Extreme Weather Variability on Hot Rocky Exoplanet 55 Cancri e Explained by Magma Temperature-Cloud Feedback}

\author[a,1]{Kaitlyn Loftus}
\author[b,1,2]{Yangcheng Luo}
\author[c]{Bowen Fan}
\author[c]{Edwin S. Kite}

\affil[a]{Climate School, Columbia University, New York, NY 10025, USA}
\affil[b]{Laboratoire de Météorologie Dynamique/Institut Pierre-Simon Laplace, Sorbonne Université, École Normale Supérieure, Université Paris Sciences et Lettres, École Polytechnique, Institut Polytechnique de Paris, Centre National de la Recherche Scientifique, Paris 75005, France}
\affil[c]{Department of the Geophysical Sciences, The University of Chicago, Chicago, IL 60637, USA}

\equalauthors{\textsuperscript{1}K.L. and Y.L. contributed equally to this work.}
\correspondingauthor{\textsuperscript{2}To whom correspondence should be addressed: yangcheng.luo@lmd.ipsl.fr}
\dates{Compiled: \today}


\begin{document}

{\parindent0pt
\infopage  
{\noindent\huge\sffamily\bfseries Abstract\par}\vspace{5mm}
Observations of the hot rocky exoplanet 55 Cancri e report significant but unexplained variability in brightness across visible and infrared bands, e.g., on sub-weekly timescales, its mid-infrared brightness temperature fluctuates by approximately 1400~K (with hundreds of Kelvin uncertainty). We propose a magma temperature-cloud feedback as a potential explanation that relies on the planet's atmosphere and surface. In this feedback, under cloud-free conditions, stellar radiation heats surface magma, releasing silicate vapor that condenses into clouds. Once formed, these clouds attenuate stellar insolation, thereby cooling the surface, reducing vapor supply, and decreasing cloudiness. A time lag between surface temperature increase and cloud formation, likely due to lagged atmospheric transport of cloud-forming vapor, enables self-sustained oscillations in surface temperature and cloudiness. These oscillations manifest as variations in both the planet's thermal emission and reflected starlight, causing variability in secondary eclipse depths across wavelengths without significantly affecting the transit depth. Using a simple model, we find that diverse planetary parameters can reproduce the observed infrared brightness variability. We also demonstrate that brightness at different wavelengths can oscillate out of phase, consistent with recent observations by the James Webb Space Telescope. Additionally, we propose that time-varying and spatially non-uniform cloud cover can result in changing amplitude and phase offset of the planet's phase curve, potentially explaining observations. Finally, we discuss observational strategies to test this proposed mechanism on 55 Cancri e. If confirmed, these observable ocean-atmosphere dynamics on exoplanets would provide valuable insights into the composition, evolution, and long-term fate of rocky planet volatiles.
\clearpage

{\noindent\huge\sffamily\bfseries Significance Statement\par}\vspace{5mm}
This study proposes a feedback between magma ocean surface temperature and reflective silicate cloud cover and shows how it could generate self-sustained oscillations on ultra-hot rocky planets with atmospheres, driving extreme weather variability. These oscillations may explain three puzzling dynamic features of the exoplanet 55 Cancri e: its varying brightness, the changing amplitude of brightness variations over an orbit, and the shifting location of the brightest point on the planet. This work provides insight into how planetary-scale ocean-cloud feedbacks can drive nonlinear behaviors like self-oscillations in climate systems, potentially inspiring research on similar phenomena on lava planets and their observational prospects. If confirmed, this study would open a new window into the composition, evolution, and fate of rocky planet volatiles.
\clearpage

}

%

{\noindent\huge\sffamily\bfseries Introduction\par}\vspace{5mm}
The super-Earth 55~Cancri~e is an extraordinary target for exoplanet observation and planetary geoscience. The planet closely orbits a Sun-like star \cite{Winn2011,Demory2011}, which heats 55 Cancri e to extreme temperatures, making its 3~$\times$~10$^{21}$~W thermal emission distinguishable from that of the host star. Its short orbital period ($<$18 hours) allows for multiple measurements of the planet’s thermal emission flux through secondary eclipse photometry or spectroscopy using space telescopes, such as Spitzer and the James Webb Space Telescope (JWST) \cite{Demory2012,Demory2016a,Demory2016b,Tamburo2018,Mercier2022,Hu2024,Patel2024}. Recent JWST observations of this super-Earth, using secondary eclipse spectroscopy, have suggested the presence of an atmosphere rich in $\text{CO}_2$ or CO \cite{Hu2024}. If confirmed, this would mark the first detection of an atmosphere surrounding a rocky exoplanet.

A notable puzzle revealed by these observations is the temporal variability of the planet's thermal emission flux, of unknown cause. This variability implies that 55~Cancri~e ``undergoes a global energy balance change more than 10,000 times greater than that from anthropogenic climate change on Earth'' \cite{Kite2021}. In the Spitzer IRAC 4.5 \textmu m channel, Demory et al.~\cite{Demory2016a} reported that the secondary eclipse depth varied between $39 \pm 25$ parts-per-million (ppm) and $212\pm46$ ppm over eight occultations within two years, corresponding to a 4.5~\textmu m brightness temperature varying between 1300$^{+270}_{-350}$ K and 2800$^{+360}_{-370}$ K. Subsequent reanalysis corroborated this variability \cite{Tamburo2018}. Recent analysis of JWST/NIRCam observations reported secondary eclipse depths in the 4.5~\textmu m band that rapidly varied from 7$\pm8.8$ ppm to 119$^{+34}_{-19}$ ppm within one week, corresponding to a 4.5~\textmu m brightness temperature varying between 870$^{+170}_{-190}$ K and 2300$^{+330}_{-190}$ K \cite{Patel2024}.

In addition to the variability in the mid-infrared (mid-IR), varying secondary eclipse depths across visible to near-IR bands have also been observed \cite{MeierValdes2022,Demory2023,Patel2024}. TESS observations in the 600–1000 nm bandpass show secondary eclipse depths varying between 0 and 15$\pm4$ ppm over two years \cite{MeierValdes2022}, CHEOPS observations in the 400–1000 nm bandpass show variations between 
0 and $166^{+22}_{-26}$ ppm over two years \cite{Demory2023}, and JWST/NIRCam observations report variations between 0 and 47$^{+21}_{-16}$ ppm at 2.1~\textmu m \cite{Patel2024}. Both reflected starlight and the planet’s thermal emission can contribute to the brightness and thus secondary eclipse depths in the visible and near-IR bands, so this visible and near-IR variability could indicate changes in both sources. The observed close-to-zero secondary eclipse depths in these bands suggest that the planet at times has a very low albedo. In contrast, observations of transit depths have not yet provided conclusive evidence of temporal variability \cite{Demory2016a,Tamburo2018,MeierValdes2022}.

Several physical mechanisms have been proposed to explain the observed variability in the secondary eclipse depth, including variability in the stellar flux induced by stochastically appearing star spots \cite{Tamburo2018}, magnetic star-planet interaction \cite{Bourrier2018a,Folsom2020}, catastrophic planetary disintegration \cite{MeierValdes2022}, an inhomogeneous dust torus around the planet \cite{MeierValdes2023}, asynchronous rotation of the planet with a hot spot on its surface \cite{Tamburo2018}, volcanic plumes temporarily elevating the altitude of the planetary photosphere \cite{Demory2016a}, and an imbalance between stochastic outgassing and atmospheric escape leading to transient atmospheres \cite{Heng2023}. Among these mechanisms, variability in the stellar flux is considered unlikely \cite{Tamburo2018} as the host star is quiet \cite{Fischer2008}. Star-planet interaction has been ruled out because the energy associated with magnetic star-planet interaction is orders of magnitude smaller than that required to induce the observed variations \cite{Morris2021}. Planetary disintegration is disfavored because an asymmetric transit shape indicative of a tail of a disintegrating planet has not been observed \cite{MeierValdes2022}. The hypothesis of a dust torus has difficulty explaining the relatively constant transit depths \cite{Heng2023}. Recent JWST observations have found that the variability in the secondary eclipse depths in two bandpasses is not correlated and thus do not support the hypothesis of a 3:2 spin-orbit resonance as the explanation \cite{Patel2024}. For the hypothesis involving transient atmospheres caused by stochastic outgassing \cite{Heng2023,Heng2024}, it is challenging for both atmospheric escape and outgassing rates to be sufficiently high to account for the variability (timescales of a week or less are reported \cite{Patel2024}).

Hu et al. \cite{Hu2024} recently speculated that atmospheric variability, tied to changes in gas composition or short-lived clouds formed from magma ocean condensates, could explain 55 Cancri e's thermal emission variability. This study proposes in greater detail an atmospheric mechanism involving reflective silicate clouds and their self-regulating effect on their own formation rate \cite[e.g.,][]{Rappaport2012,Booth2023,Bromley2023} to explain the observed variability in the infrared and visible brightness. We suggest that reflective clouds, formed from silicate vapor evaporated from a magma ocean, can attenuate stellar radiation reaching the surface, leading to surface cooling and a slowdown of the evaporation of silicates, thereby suppressing further cloud formation. We employ a simple model to demonstrate that this feedback, coupled with a delayed response of cloudiness to surface temperature change due to various possible processes, could result in self-sustained oscillations in the thermal emission flux and reflected starlight from the planet. Building on plausible atmospheric circulation patterns simulated in previous studies \cite[e.g.,][]{Hammond2017}, we further demonstrate that the time-varying, spatially non-uniform cloud coverage resulting from the proposed mechanism could produce variations in both the amplitude and phase offset of the planet’s phase curve, potentially explaining observational data \cite{Demory2016b,Sulis2019,Morris2021,Mercier2022,MeierValdes2023}.

This study builds on previous modeling works on 55 Cancri e, which suggested that clouds composed of refractory materials, such as magnesium silicates and silicon oxides, could form and even exhibit high opacity \cite{Mahapatra2017,Hammond2017}. This study also draws on the ideas from previous modeling works for clouds on brown dwarfs and giant exoplanets, which proposed that radiative cloud feedbacks could drive observed oscillations in these objects' brightness temperatures \cite{Tan2019,Tan2021a,Tan2021b}.

\vspace{6mm}{\noindent\huge\sffamily\bfseries Results\par}\vspace{5mm}
\subsection*{Self-Oscillations in Surface Temperature and Cloudiness}

\begin{SCfigure*}[\sidecaptionrelwidth][t!]
\centering
\includegraphics[width=11.4cm]{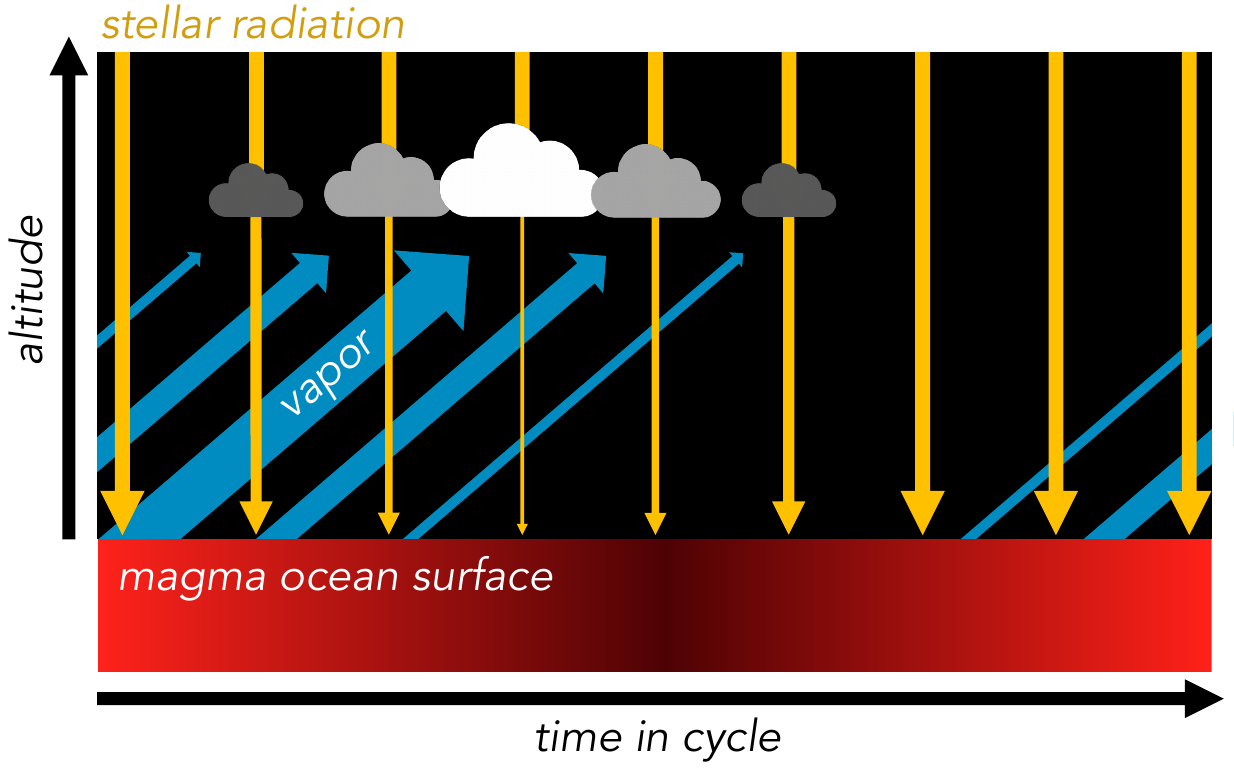}
\caption{Key elements in the magma temperature-cloud feedback that drives the oscillations in cloudiness and surface temperature. Changes in the color of the magma ocean surface indicate variations in surface temperature (brighter denotes higher temperature). Changes in the brightness of the clouds represent variations in the shortwave opacity of the cloud (brighter denotes higher optical depth). The thicknesses of downward yellow rays indicate shortwave radiation before and after being attenuated by the clouds. The thicknesses of the blue arrows indicate ascending vapor fluxes.}\label{fig:schematic}
\end{SCfigure*}

\begin{table}[t!]
\centering
\caption{Estimated bounds for the model parameters that span the parameter space}
\label{tab:varied_param}
\begin{tabular}{cccc}
Parameter & Lower bound & Upper bound & Unit \\
\midrule
$\Pi_1$ & $10^{-3.20}$ & $10^{2.41}$ & Pa$^{-1}$ \\
$\Pi_2$ & $10^{-7.34}$ & $10^{3.71}$ & dimensionless \\
$\Pi_3$ & $10^{0.439}$ & $10^{6.26}$ & kg s$^{-3}$ K$^{-1}$ \\
$\Delta T_{\text{cloud}}$ & 0 & 350 & K \\
$\beta$ & $\mathrm{10^{-3}}$ & 4 & dimensionless \\
\bottomrule
\end{tabular}

\addtabletext{Note: $\beta=0$ is also explored.}
\end{table}

We develop a box model to capture the dynamics of the magma temperature-cloud feedback, with its key components illustrated in Fig. \ref{fig:schematic}. The model tracks temporal changes in the surface temperature of the substellar magma ocean ($T_{\text{surf}}$) and the optical depth of clouds in the substellar region ($\tau_{\text{SW}}$ in the shortwave and $\tau_{\text{LW}}$ in the longwave). These variables influence each other in a feedback loop. On one hand, the surface temperature governs the supply of silicate vapor and thus the cloud production rate, following the Clausius-Clapeyron relationship. A lag parameter $d$ captures the time delay in cloud formation due to the atmospheric transport time of vapor from the surface to the cloud zone. Without this time lag, the system does not produce variability (see the Supporting Information). Cloud dissipation, meanwhile, occurs as particles sediment out of the atmosphere. On the other hand, cloud optical depth affects surface temperature by modulating the surface radiative energy budget. Shortwave cloud opacity scatters incoming stellar shortwave radiation back to space, reducing surface heating. Longwave cloud opacity leads to a greenhouse effect that warms the surface: clouds absorb part of surface-emitted longwave radiation and re-emit it back towards the surface. Together, these interactions allow the model to explore how surface temperature and cloudiness mutually influence each other and co-vary with time. Given the uncertain conditions of the atmosphere and magma ocean on the planet, we investigate the system’s behavior across a range of assumed values for cloud productivity ($\Pi_1$ in Table \ref{tab:varied_param}), cloud removal efficiency ($\Pi_2$), thermal inertia of the surface magma ocean ($\Pi_3$), cloud thickness (with the difference in downward and upward cloud emission temperatures $\Delta T_{\text{cloud}}$ as a proxy), and the strength of the cloud greenhouse effect ($\beta$) (see Materials and Methods). The parameter ranges explored arise from physical arguments (Materials and Methods, Table \ref{tab:S1}), which are consistent with mass and radiative energy conservation. We note that to reduce the number of explored parameters, we nondimensionalize time using the lag $d$; therefore, we report all results in time units of $d$. 

\begin{figure*}[t!]
\centering
\includegraphics[width=17.8cm]{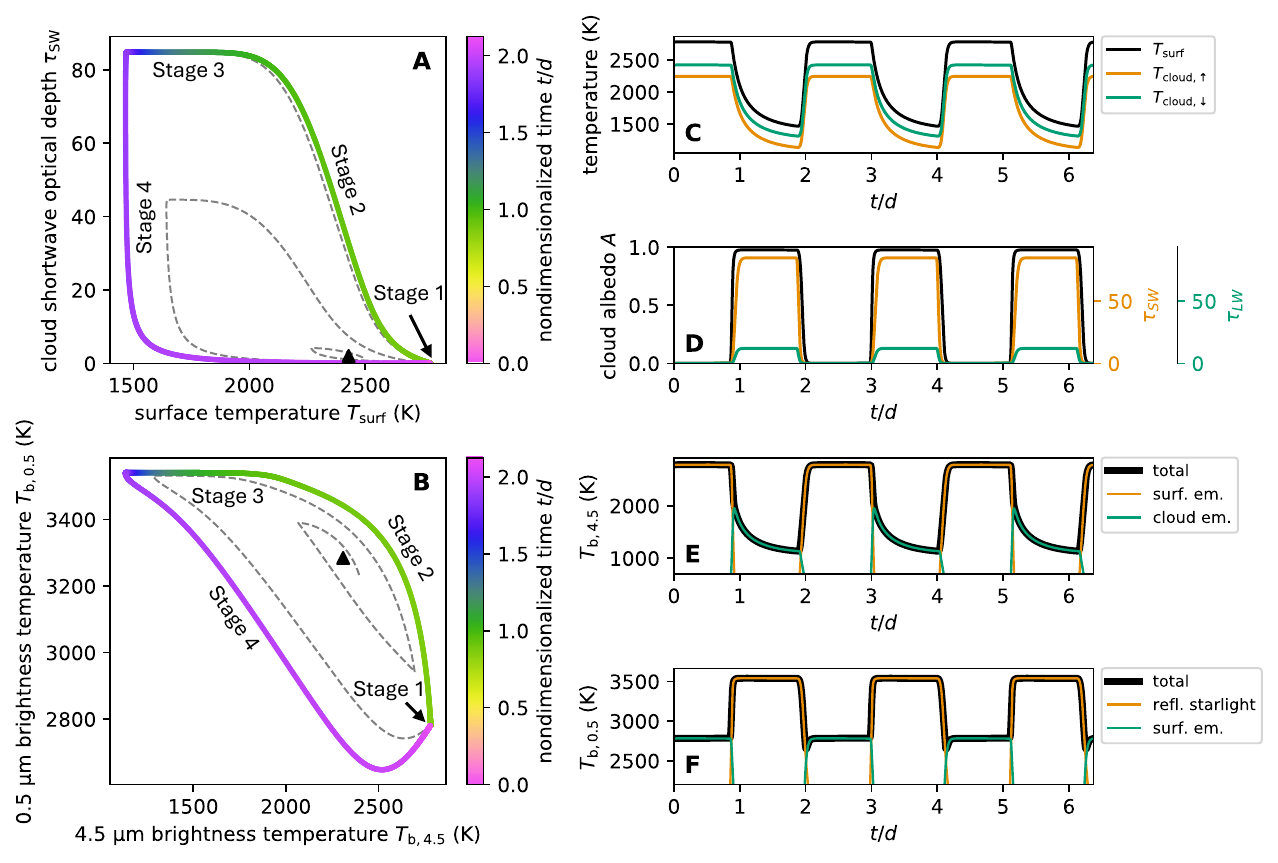}
\caption{An example of simulated limit-cycle oscillations in surface temperature ($T_{\text{surf}}$) and shortwave cloud optical depth ($\tau_{\text{SW}}$) in the substellar region, which manifest as oscillations in substellar brightness temperature at 4.5~\textmu m in the mid-IR band ($T_{\text{b,4.5}}$) and at 0.5 \textmu m in the visible band ($T_{\text{b,0.5}}$). ($A$) Phase diagram of $T_{\text{surf}}$ and $\tau_{\text{SW}}$, where the trajectory of the system’s state vector (dashed curve) converges towards a closed loop (colored curve) while spiraling counter-clockwise, indicating a stable limit cycle. Colors represent nondimensionalized time (time $t$ normalized by the lag $d$) over the course of one cycle period. The black triangle marks the unstable fixed point. ($B$) Similar to ($A$), but for the 4.5~\textmu m brightness temperature $T_{\text{b,4.5}}$ and the 0.5~\textmu m brightness temperature $T_{\text{b,0.5}}$. ($C$) Time series of surface temperature $T_{\text{surf}}$ (black), cloud upward emission temperature $T_{\text{cloud}\uparrow}$ (orange), and cloud downward emission temperature $T_{\text{cloud}\downarrow}$ (green) over three cycles. Time is again nondimensionalized using $d$. ($D$) Similar to ($C$), but showing shortwave cloud albedo $A$ (left y-axis, black), shortwave cloud optical depth $\tau_{\text{SW}}$ (right y-axis, orange), and longwave cloud optical depth $\tau_{\text{LW}}$ (right y-axis, green). ($E$) Time series of brightness temperature at 4.5 µm. The black curve represents the brightness temperature resulting from the combined thermal emission from the surface and the cloud. The orange curve shows only surface emission penetrating the cloud, and the green curve shows only cloud emission. ($F$) Similar to ($E$), but for the brightness temperature at 0.5~\textmu m. The black curve represents the brightness temperature resulting from the combined reflected starlight and surface thermal emission, while the orange and green curves represent reflected starlight and surface thermal emission, respectively. Model parameters: $\Pi_1=9.47$ $\mathrm{Pa}^{-1}$, $\Pi_2=74.2$, $\Pi_3=115$ $\mathrm{kg}$ $\mathrm{s}^{-3}$ $\mathrm{K}^{-1}$, $\Delta T_{\text{cloud}}=177$ $\mathrm{K}$, and $\beta=0.143$.}\label{fig:limitcycleex}
\end{figure*}

Limit-cycle oscillations in surface temperature ($T_{\text{surf}}$) and shortwave cloud optical depth ($\tau_{\text{SW}}$) occur within certain regions of the parameter space defined in Table \ref{tab:varied_param}. Figure \ref{fig:limitcycleex} illustrates an example. In the phase diagram for $T_{\text{surf}}$ and $\tau_{\text{SW}}$ (Fig. \ref{fig:limitcycleex}A), the limit cycle is represented by a closed loop. If the system’s state starts inside the loop, it spirals outward until it converges on the loop; if it starts outside, it spirals inward towards the loop. A state initiated precisely on the loop remains on the loop indefinitely. These behaviors indicate that the closed loop forms a stable limit cycle, making the system’s oscillations an intrinsic and resilient feature of its dynamics. Thus, the oscillatory behavior is robust against perturbations: while one-time disturbances may temporarily alter the amplitude or the period of oscillations, the system returns to its intrinsic oscillatory state after the perturbations stop. There is a fixed point inside the loop. The system’s state remains unchanged if initiated at this point, but any deviation from the fixed point will amplify, driving the system towards the limit cycle. This indicates that the fixed point is unstable. Given external perturbations in the real world, this instability suggests that no steady state can exist in these regions of the parameter space.

As shown in Fig. \ref{fig:limitcycleex}, a typical limit cycle can be divided into four stages:

Stage 1: The system is nearly cloud-free, with the surface receiving unattenuated stellar radiation. The surface stays at its highest temperature to balance the heating from stellar radiation with cooling through thermal emission. This hot-surface, cloud-free state typically lasts for a duration between $d$ and $2d$, appearing as a point in Fig. \ref{fig:limitcycleex}A and nearly horizontal lines in Figs. \ref{fig:limitcycleex}C and \ref{fig:limitcycleex}D. During this stage, the evaporation of silicates, which supplies cloud-forming vapor, is strong, but the vapor has not yet reached the colder, cloud-forming regions due to lagged atmospheric transport.

Stage 2: This stage begins when the vapor reaches the cold regions where the vapor condenses, ultimately leading to an increase in cloudiness over the substellar region. This stage corresponds to the sudden rise in cloud optical depth in Figs. \ref{fig:limitcycleex}A and \ref{fig:limitcycleex}D. The increased cloud albedo reflects a significant portion of stellar radiation, reducing the energy absorbed by the surface and cooling the surface.

Stage 3: The cloud optical depth remains at its maximum (Figs. \ref{fig:limitcycleex}A and \ref{fig:limitcycleex}D), minimizing the amount of stellar radiation reaching the surface. Surface cooling persists as a result. Following the Clausius-Clapeyron relationship, the decrease in surface temperature lowers the saturated vapor pressure, significantly reducing vapor supply from evaporation. 
However, the cloud does not immediately respond due to the time lag in vapor transport from the substellar magma ocean surface to the cloud formation zone. This stage typically lasts longer than $d$.

Stage 4: The surface cooling effect eventually reaches the cloud-forming regions, causing a rapid decline in cloud formation. As clouds fade out by particle sedimentation, the atmosphere returns to a clear-sky condition. With the clouds dissipating, more stellar radiation reaches the surface, which warms up. This stage ends when the clouds have almost disappeared and the surface temperature has risen sufficiently to balance incoming shortwave radiation through thermal emission under clear skies, marking the start of the next cycle.

The longwave cloud optical depth is assumed to be proportional to the shortwave optical depth (Eq. \ref{eq:taulinearity}), so the longwave cloud optical depth also increases during Stage 2 (Fig. \ref{fig:limitcycleex}D), enhancing the greenhouse effect that warms the surface. However, this warming effect is transient: as the surface cools, thermal emission from the surface, which provides the energy sustaining the cloud’s temperature, also declines, so the cloud cools with the cooling surface (Fig. \ref{fig:limitcycleex}C), reducing surface warming from cloud longwave radiation. In most cases, surface heating by cloud thermal emission is relatively minor compared to the heating from absorbed stellar radiation and the cooling from surface thermal radiation throughout the cycle.

\subsection*{Resulting Oscillations in Planetary Brightness}

The oscillations in surface temperature and cloud optical depth drive corresponding fluctuations in brightness temperature in both mid-IR and visible bands (Figs. \ref{fig:limitcycleex}B, \ref{fig:limitcycleex}E, and \ref{fig:limitcycleex}F). We focus on the brightness temperatures at 4.5~\textmu m for the mid-IR ($T_{b,4.5}$) and at 0.5~\textmu m for the visible band ($T_{b,0.5}$) to illustrate these processes.

Stage 1: During this stage, the surface is at its hottest state, and the atmosphere is nearly cloud-free. Almost all of the strong thermal emission from the surface escapes directly to space, resulting in $T_{b,4.5}$ being at its peak, closely matching the surface temperature (about 2780 K, Figs. \ref{fig:limitcycleex}B and \ref{fig:limitcycleex}E). In contrast, $T_{b,0.5}$ is near its minimum (about 2780 K) because, without reflective clouds, no reflected starlight contributes to the visible brightness, so $T_{b,0.5}$ also aligns with the surface temperature (Figs. \ref{fig:limitcycleex}B and \ref{fig:limitcycleex}F).

Stage 2: As cloud optical depth increases rapidly, the surface temperature plummets. This leads to a quick decline in $T_{b,4.5}$ as the mid-IR emission level rises from the surface to the cloud-top (Fig. \ref{fig:limitcycleex}E). Simultaneously, $T_{b,0.5}$ rises due to increased cloud albedo reflecting more starlight (Fig. \ref{fig:limitcycleex}F).

Stage 3: The cloud optical depth remains high while surface and cloud temperatures continue to fall. This causes a continued decline in $T_{b,4.5}$, as it is now dominated by thermal emission from the cloud (Fig. \ref{fig:limitcycleex}E). In contrast, $T_{b,0.5}$ stays at its peak due to the constant stellar radiation and stable cloud albedo (Fig. \ref{fig:limitcycleex}F).

Stage 4: Cloud optical depth decreases rapidly, exposing the surface, and surface temperature begins to rise again, dominating $T_{b,4.5}$. As the surface re-heats from stellar radiation, $T_{b,4.5}$ increases quickly (Fig. \ref{fig:limitcycleex}E). Meanwhile, $T_{b,0.5}$ drops rapidly as the reflective cloud disperses, reducing the contribution from reflected starlight (Fig. \ref{fig:limitcycleex}F). 

Broadly, our model predicts that $T_{b,0.5}$ tends to increase when $T_{b,4.5}$ decreases and vice versa; however, for many parameter cases, $T_{b,4.5}$ and $T_{b,0.5}$ are not exactly anti-correlated and instead oscillate out of phase (Fig. \ref{fig:limitcycleex}B). During the end of Stage 4, $T_{b,0.5}$ can decrease to a minimum and begin to rise again while $T_{b,4.5}$ is increasing, because the surface temperature does not rise fast enough to compensate for the reduced reflected starlight. Although the dip and rebound of $T_{b,0.5}$ in the example case in Fig. \ref{fig:limitcycleex}B are transient, they are more pronounced (10–20\% of the oscillation period in duration) in other cases (e.g., Fig. \ref{fig:s1}). During Stage 3, $T_{b,0.5}$ can remain at its maximum (or near it) while $T_{b,4.5}$ substantially decreases: as cloud cover increases, $T_{b,0.5}$ reaches its peak due to maximum cloud albedo, while $T_{b,4.5}$ continues to fall as the cloud and surface cool. These out-of-phase behaviors are found across a broad parameter space where oscillations occur, though the exact phase difference between $T_{b,4.5}$ and $T_{b,0.5}$ oscillations depends on specific model parameters. (Figures \ref{fig:s1}–\ref{fig:S2} show example phase diagrams of $T_{b,4.5}$ and $T_{b,0.5}$ with different parameter values from Fig. \ref{fig:limitcycleex}.) The near-IR brightness temperature, observed at 2.1 µm by JWST \cite{Patel2024}, lies between the mid-IR and visible bands and is also expected to oscillate out of phase with visible and mid-IR brightness temperatures. However, detailed modeling based on wavelength-dependent optical properties of silicate clouds is out of the scope of this study.

\subsection*{Influence of Model Parameters on Model-Observation Consistency}

\begin{SCfigure*}[\sidecaptionrelwidth][t!]
\centering
\includegraphics[width=11.4cm]{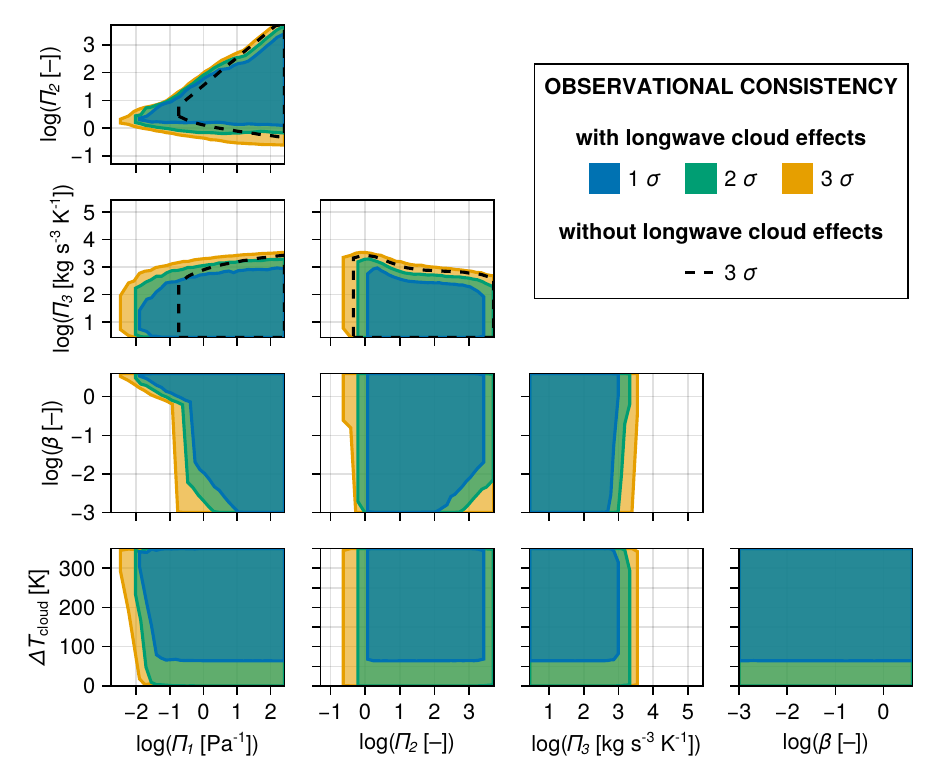}
\caption{Parameter ranges capable of simultaneously reproducing the upper and lower bounds of Spitzer and JWST 4.5 \textmu m brightness temperature observations within one, two, and three times the measurement uncertainty ($\sigma$). The shading and solid contours show the model with longwave cloud effects ($10^{-3} < \beta < 4$). Different colors indicate different levels of consistency. The dashed contours in the first two rows show the model without longwave cloud effects ($\beta=0$) at $3\sigma$ model-observation consistency. No $2\sigma$ consistency can be achieved without longwave cloud effects. Each panel shows a two-dimensional (2D) projection of the 5D (for $10^{-3}<\beta<4$) or 3D (for $\beta=0$) parameter space defined in Table \ref{tab:varied_param}. A location in each 2D projection is classified into a given consistency tier if, for the two parameters plotted, there exist values of the other parameters (within their respective bounds) that can produce oscillations consistent with observations at that consistency tier. Consistency between the model and observations is achieved across a substantial subset of the parameter space.}\label{fig:parambounds}
\end{SCfigure*}

Figure \ref{fig:parambounds} illustrates regions of the parameter space (Table \ref{tab:varied_param}) where limit-cycle oscillations can reproduce the observed highest and lowest 4.5~\textmu m brightness temperatures \cite{Demory2016a,Patel2024}. A substantial range of model parameters is capable of reproducing the observed magnitude of the 4.5~\textmu m brightness temperature variability. Favorable conditions for reproducing the observed oscillations include a large $\Pi_1$, an intermediate $\Pi_2$, a small $\Pi_3$, and a large $\Delta T_{\text{cloud}}$ in the presence of longwave cloud effects ($\beta > 0$).

Model-observation consistency requires that $\Pi_1$, which governs vapor supply and cloud formation, is not too small, because insufficient cloud formation fails to significantly reduce surface heating by stellar radiation in Stage 2. An intermediate $\Pi_2$, which corresponds to an intermediate cloud removal rate, is also required. If $\Pi_2$ is too small, clouds persist too long, damping oscillations in cloud optical depth and reducing their amplitude. Conversely, if $\Pi_2$ is too large, limited cloudiness resulting from rapid cloud removal prevents sufficient surface cooling, making it hard to match the observed lowest 4.5 \textmu m brightness temperature. A small $\Pi_3$ helps sustain large-amplitude oscillations by allowing fast surface temperature response to imbalances in the surface energy budget. If $\Pi_3$ is too large, surface temperature oscillations are dampened, resulting in small oscillation amplitude.

The model parameter $\beta$, which represents the ratio of longwave to shortwave cloud optical depth, affects the strength of the cloud greenhouse effect. $\beta$ generally has a small impact as long as $\beta > 10^{-3}$, except when $\Pi_1$ is small or $\Pi_2$ is large. In such cases, limited cloud formation or rapid cloud removal leads to limited cloudiness, so a larger $\beta$—indicating a higher capability of a single cloud particle to absorb longwave radiation—allows stronger absorption of surface radiation by cloud, thereby enhancing the greenhouse effect and helping to reduce $T_{b,4.5}$ to the observed values \cite{Patel2024}. Notably, in the absence of longwave cloud absorptivity and emissivity ($\beta=0$), surface thermal emission alone can reproduce the observations within $3\sigma$ in some regions of the parameter space. However, the solidus temperature of the magma imposes a lower limit on the surface temperature, preventing $2\sigma$ consistency with the lowest observed $T_{\text{b,4.5}}$ from ref. \cite{Patel2024}. Achieving $2\sigma$ and $1\sigma$ model-observation consistency requires the cloud to absorb thermal radiation from the surface and re-emit it at a lower temperature. In some rare cases, we find $\beta>1$ can lead to erratic variability; we conservatively exclude these erratic cases from model-observation consistency tests (affecting less than 0.2\% of cases) and discuss them in detail in the Supporting Information (Fig. \ref{fig:erratic_example}–\ref{fig:betaevo}). 

With longwave cloud absorptivity and emissivity ($\beta > 0$), the parameter $\Delta T_{\text{cloud}}$, which represents the difference between the cloud’s downward and upward emission temperatures (largely influenced by cloud layer thickness), also impacts the mid-IR brightness. If $\Delta T_{\text{cloud}}$ is too small, the cloud-top temperature remains too high, meaning that even if all surface radiation is absorbed by the cloud, the cloud’s own thermal emission could result in a $T_{b,4.5}$ significantly higher than the observed minimum.

Figure \ref{fig:parambounds} also shows that a smaller $\Pi_2$ can offset the effects of a smaller $\Pi_1$. When vapor supply and cloud formation are inefficient (small $\Pi_1$), a slower cloud particle sedimentation rate or longer cloud residence time (small $\Pi_2$) can sustain higher cloudiness, which helps sufficiently cool the surface. However, as previously noted, $\Pi_2$ cannot be too small, as long-lived clouds would dampen the oscillations.

The oscillation period ranges from 2$d$ to 11.5$d$ for oscillations consistent with observations within 3$\sigma$ and 2$\sigma$, and from 2$d$ to 9$d$ for those within 1$\sigma$. We discuss how these periods relate to the observed time scales of variability in the next section. Among the model parameters, $\Pi_2$ has the strongest effect on the oscillation period, with smaller values of $\Pi_2$ resulting in longer periods; a smaller $\Pi_2$ increases the cloud residence time, extending the duration of Stage 4 (during which cloud particle sedimentation occurs). Similarly, $\Pi_3$ also affects the oscillation period, with larger $\Pi_3$ leading to longer periods. A larger $\Pi_3$ slows the rate of surface temperature change, lengthening Stages 2–4.

\vspace{6mm}{\noindent\huge\sffamily\bfseries Discussion\par}\vspace{5mm}
\subsection*{Observational Tests}

We have shown that the magma temperature-cloud feedback can induce oscillations in the brightness temperature of 55 Cancri e (Figs. \ref{fig:limitcycleex}B, \ref{fig:limitcycleex}E, and \ref{fig:limitcycleex}F), potentially explaining observational data. In the mid-IR (e.g., 4.5~µm), the highest brightness temperature in the substellar region can reach $\sim$2780 K, corresponding to the maximum surface temperature after an extended period of clear-sky conditions. The lowest brightness temperature of the substellar region can be as low as $\sim$960 K, which corresponds to the lowest cloud top emission temperature resulting from the lowest surface temperature (set by the solidus temperature of magma) after an extended cooling period with maximum cloudiness. Since the substellar region is likely the hottest area on the planet, these modeled substellar values are likely upper bounds on the full disk-averaged brightness temperature of the planetary hemisphere facing the star (i.e., the ``dayside''), which is retrieved from observations of secondary eclipse depth. We assume 55 Cancri e is tidally locked \citep{Patel2024}, which means that the same ``dayside'' hemisphere (with the substellar point in the center) is always facing the star and the other ``nightside'' hemisphere is always facing away from the star.

At 0.5~\textmu m in the visible band, the highest brightness temperature in the substellar region can reach $\sim$3500 K, corresponding to nearly full reflection of starlight at maximum cloud reflectivity. The lowest brightness temperature in the substellar region corresponds to the lowest surface temperature immediately after cloud dispersion exposes the surface, which we find to be higher than $\sim$1400~K for models that align with the mid-IR observations. As reflected starlight could make a significant contribution to the planet's brightness in the shortwave and the associated secondary eclipse depth, estimating the oscillation amplitudes of these two quantities requires knowledge about the horizontal distribution of reflective cloud over the dayside and wavelength-dependent cloud particle optical properties, including the intensity of backscattering. As our simple model does not predict these effects, it is challenging to make predictions for the planet's shortwave brightness temperature and secondary eclipse depths. Higher dimensional simulations that couple silicate cloud evolution with a surface magma ocean are needed to better constrain cloud spatial and temperature distributions as well as cloud particle composition and shape. Because small changes in composition can lead to large changes in particle optical properties \cite{Zeidler2011}, laboratory experiments that measure composition and optical properties of particles formed from vaporized magma are highly desirable \cite[e.g.,][]{He2024}. Such future modeling and experimental efforts will enable more detailed predictions of the magma temperature-cloud feedback's implications for shortwave brightness temperature and secondary eclipse depth.

Still, within the framework of our proposed mechanism, we can estimate an upper bound for the secondary eclipse depth in the shortwave where reflected starlight dominantly contributes. If clouds scatter starlight like a Lambertian surface, the geometric albedo due to cloud reflectivity cannot exceed 1, corresponding to a secondary eclipse depth less than $\sim$30~ppm at all wavelengths in the shortwave. Our model predicts a negligible contribution from the planet's thermal emission to the visible-band brightness during time periods with maximum cloudiness. Therefore, our proposed mechanism cannot produce broad-band shortwave secondary eclipse depths higher than $\sim$30~ppm, unless silicate clouds backscatter more strongly than a Lambertian surface. The majority of the observed secondary eclipse depths in refs. \cite{MeierValdes2022} and \cite{Demory2023} are compatible with this upper bound within twice observational uncertainty, but the high outlier ($>$ 50~ppm) eclipse depths reported in ref. \cite{Demory2023}, if confirmed, could pose a challenge to cloud and surface temperature variability as the sole explanation for the observed secondary eclipse depth variability. However, we note that our cloud temperature model might have neglected important heat sources, such as latent heat release during condensation, so the cloud top temperature is likely higher than predicted by our model. The resulting stronger thermal emission from the cloud could enhance the planet's brightness in CHEOPS's band pass (400~nm--1~\textmu m), especially in the near-IR, if cloud emissivity dominates over reflectivity there.

Over six days with JWST/NIRCam, Patel et al. \cite{Patel2024} observed a minimum near-IR brightness temperature while the mid-IR brightness temperature increased (Fig. \ref{fig:JWST}). Our model predicts such out-of-phase oscillations in brightness temperature across different spectral bands. The minimum brightness temperature in the visible band can occur while the brightness temperature in the mid-IR band increases, and the maximum visible brightness temperature tends to occur during the decreasing phase of the mid-IR brightness temperature (e.g., Fig. \ref{fig:limitcycleex}B, Fig. \ref{fig:s1}). Under our proposed mechanism, we also expect the brightness temperature in the near-IR band to oscillate out of phase with visible-band and mid-IR brightness temperature, as observed by Patel et al. \cite{Patel2024}, but detailed modeling requires knowledge about wavelength-dependent emissivity and reflectivity of the clouds.

The oscillation period predicted by our model is typically several times longer than the lag between the surface temperature maximum and the subsequent increase in cloudiness. This lag could be as short as the vertical transport timescale from the magma ocean surface to the cloud condensation level, likely on an hourly timescale. In such a case, the model’s oscillation period would be short, and brightness temperature variations could be partially smoothed out during the 1.6-hour secondary eclipse. However, in a scenario where ascending vapor is sheared eastward by strong westerly winds, delaying cloud formation over the substellar region, the lag could extend to about ten days. This estimate comes from dividing the planet’s size by the lowest zonal wind speeds \cite{Hammond2017}, as clouds forming on the eastern dayside would need to travel around the planet before covering the substellar region. Such a scenario would require an extended cloud residence time compared to the lag, which corresponds to a relatively small $\Pi_2$. Abundant background volatile atmospheric gases such as $\mathrm{N_2}$, CO, and $\mathrm{CO_2}$ are likely necessary. The observed sub-weekly-timescale brightness temperature variability \cite{Patel2024} suggests a timescale that lies between the oscillation periods derived from these bounds on the lag. Other processes, such as the breaking of a near-surface convective inhibition layer at the onset of surface warming \cite{Colby1984} or the nucleation of cloud particles \cite{Lee2018,Helling2019}, are also expected to influence the lag and hence the oscillation period.

If clouds have a sufficiently long residence time, they could be transported to the nightside, allowing the oscillating cloudiness over the substellar region from our model to manifest as oscillating cloud coverage on the nightside as well. If these clouds have infrared opacity, the oscillating cloudiness on the nightside could lead to fluctuations in the nightside brightness temperature, if nightside cloud temperature differs from nightside surface temperature. Such variations could potentially be detected by future JWST observations of the planet’s phase curve.

\begin{figure*}[t!]
\centering
\includegraphics[width=17.8cm]{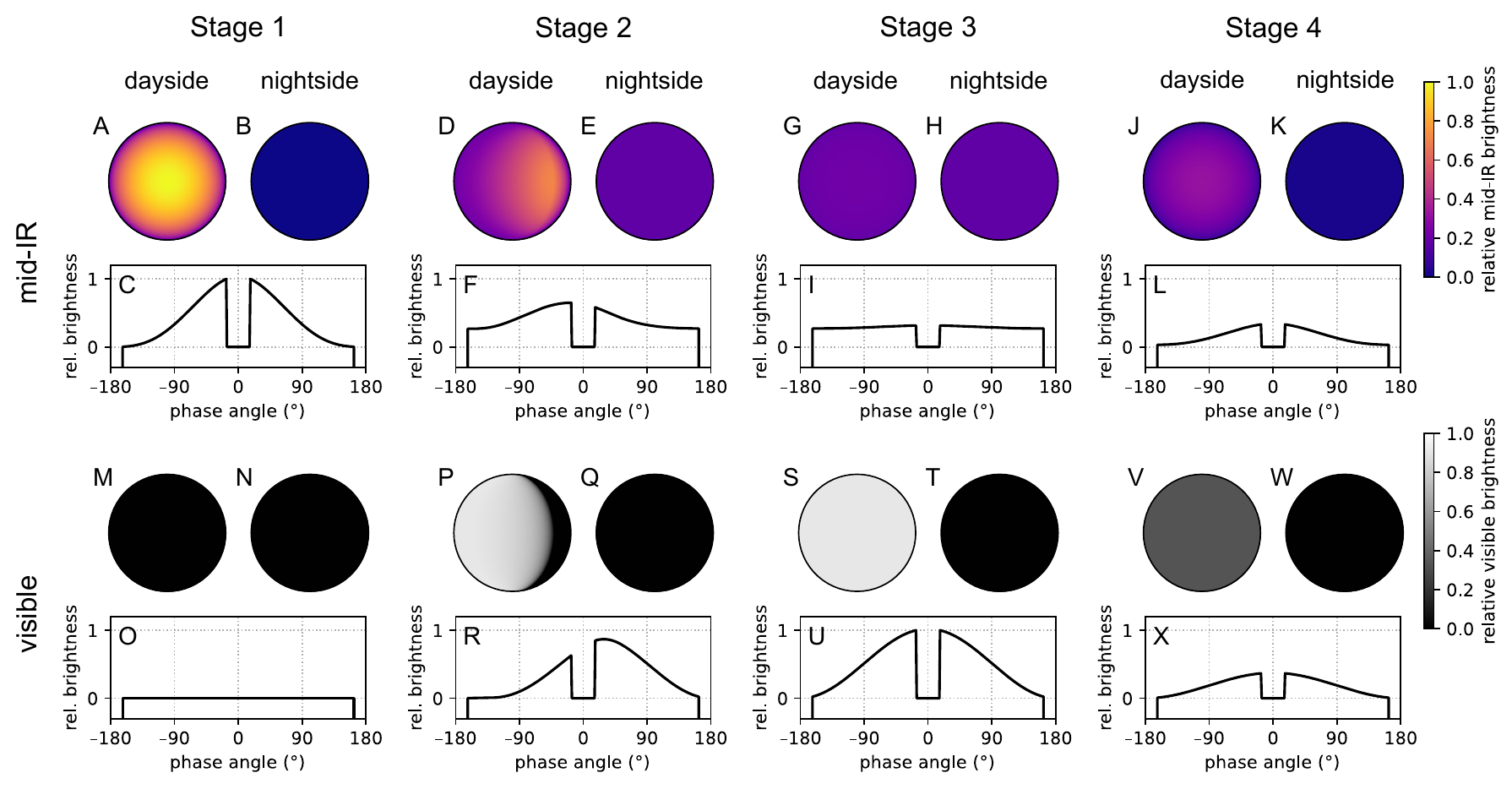}
\caption{Variable, non-uniform cloud cover can produce fluctuations in both the magnitude and phase offset of planetary phase curves across shortwave and longwave channels. Rows 1 and 3 illustrate the brightness distributions on the dayside (i.e., the hemisphere facing the star) and nightside (i.e., the hemisphere facing away from the star) of the planet, in the mid-infrared (mid-IR) and visible wavelengths, respectively. Rows 2 and 4 display the phase curves for the mid-IR and visible channels, representing how the brightness of the entire planetary disk, as seen by the observer, changes with phase angle (the orbital angle covered by the planet since the secondary eclipse). The dips around 0$^\circ$ phase angle correspond to secondary eclipses, when the planet is hidden behind the star. The dips around 180$^\circ$ phase angle corresponds to primary transits, when part of the stellar disk is behind the planet. The four columns correspond to the four stages of oscillation shown in Fig. \ref{fig:limitcycleex}, characterized by the following: no cloud cover (Stage 1), increasing cloudiness with expanding cloud coverage (Stage 2), full cloud cover (Stage 3), and diminishing cloudiness (Stage 4). The mid-IR phase curves are normalized to their maximum values for comparison, as are the visible phase curves.}\label{fig:phasecurve}
\end{figure*}

Additionally, clouds transported from the nightside to the dayside by westerlies could lead to increased cloud coverage and optical depth over the western dayside, due to cloud sedimentation and re-evaporation during further eastward transport over the dayside \citep{Powell2018,Ehrenreich2020,Coulombe2025}. This increased cloud coverage on the western dayside would result in higher brightness temperatures in the visible band due to enhanced reflection of starlight (Fig. \ref{fig:phasecurve}P–R), as reported in ref. \cite{Morris2021}. If the clouds have infrared opacity, they would absorb surface thermal radiation in the mid-IR band, leading to lower brightness temperatures on the western dayside and creating an apparent eastward ``hot spot shift,'' (Fig. \ref{fig:phasecurve}D–F), as reported in ref. \cite{Demory2016b}. These effects could cause asymmetry in the planet’s phase curve (Figs. \ref{fig:phasecurve}F, \ref{fig:phasecurve}R). If such asymmetry exists, it is most likely observable during Stage 2 in the oscillation cycle. However, it is important to note that starlight reflected by asymmetrically distributed clouds on the dayside cannot produce maximum brightness in the visible band near primary transit. Therefore, if confirmed, the peaks in visible-band brightness near primary transit reported in refs. \cite{Sulis2019,MeierValdes2023} could challenge the idea that time-varying, non-uniform cloud cover on the dayside alone explains the observed variations in the amplitude and phase offset of the planet’s visible phase curves.

In summary, our model is consistent with phase curve asymmetry of varying intensity in both the visible and mid-IR bands: the highest brightness temperature in the visible band would predominantly shift westward, while in the mid-IR band, it would shift eastward. (However, there may be instances when the visible hot spot shifts eastward and the mid-IR hot spot shifts westward, occurring just after clouds begin to form on the eastern dayside but have yet to be transported to the western dayside after their journey to the nightside.) Furthermore, the westward shift of the ``hot spot'' in the visible band and the eastward shift of the ``hot spot'' in the mid-IR band may occur simultaneously, both corresponding to higher cloud coverage over the western dayside than the eastern dayside. These phenomena may also coincide with an anomaly of nightside brightness temperature in the mid-IR, due to a cloud greenhouse effect over the nightside (Figs. \ref{fig:phasecurve}B, \ref{fig:phasecurve}E, \ref{fig:phasecurve}H, and \ref{fig:phasecurve}K). Whether this anomaly of nightside mid-IR brightness temperature is positive or negative may mainly depend on whether the nightside cloud temperature is higher or lower than the surface temperature (assumed higher in Fig. \ref{fig:phasecurve}).

The transit depth variations due to varying cloudiness are only marginally observable. The altitude of the cloud tops could extend several times the atmospheric scale height, which is estimated to be on the order of 50 km, assuming an atmosphere dominated by $\mathrm{N_2}$, CO, or $\mathrm{CO_2}$. This is comparable to the 1$\sigma$ observational uncertainty in the planet's radius, $\sim$150 km \cite{MeierValdes2022}.

Future multiband observations of 55 Cancri e’s phase curve and secondary eclipse depths will provide valuable opportunities to test the hypotheses presented in this study. Additionally, detecting spectral features of silicate vapor or silicate clouds \cite{Wakeford2015,Gao2020,Burningham2021,Grant2023,Dyrek2024} will further support this hypothesis.

\subsection*{Model Limitations}

We present a simple model of a magma temperature-cloud feedback that can explain the observed brightness temperature oscillations on 55 Cancri e. Many complex processes, (e.g., atmospheric dynamics, magma ocean dynamics, cloud microphysics)  which can be represented in more sophisticated, higher-dimensional models, have been omitted or simplified.

The model simplifies the vapor transport from the substellar magma ocean surface, where silicate-derived vapor is supplied to the atmosphere through partial vaporization of magma, to the cloud-forming region. This process is treated as vertical transport. In reality, however, strong horizontal winds may play a significant role in the vapor transport, as mentioned in Observational Tests. Earlier works predict wind speeds ranging from 0.2~$\mathrm{km}$~$\mathrm{s}^{-1}$ to 2~$\mathrm{km}$~$\mathrm{s}^{-1}$ \cite{Hammond2017,Nguyen2020}. According to general circulation models (GCMs), these strong winds likely shear the ascending motion above the substellar region eastward, causing clouds to form primarily near the eastern terminator of the dayside, where temperature is significantly lower than the substellar region. In this scenario, clouds would no longer immediately affect the stellar radiation reaching the surface after their formation. Instead, they may be transported to the nightside by westerlies and then return to the dayside across the western terminator, after which they could eventually reach the substellar region and attenuate incoming stellar radiation.

Even within the simplified vertical transport framework, additional assumptions are made. The model characterizes the ascent of vapor from the magma ocean surface to the cloud base using an ascending velocity that is proportional to the atmospheric scale height and inversely proportional to a constant vertical transport timescale. In reality, convective updrafts are influenced by other factors such as surface heating, atmospheric stability, large-scale forcings, and turbulent mixing—processes that can only be treated in high-resolution higher-dimensional models. For example, the transport timescale $d$ is not constant; during surface cooling, increased atmospheric stability can inhibit convective vapor transport. Conversely, during surface heating, the transport timescale may rapidly shorten as convection resumes \cite[e.g.,][]{Colby1984}. Thus, the assumption of a constant transport timescale likely produces oscillations that are smoother than those expected in reality.

In order to generate cloud radiative feedbacks, we assume that cloud particles nucleate and grow to sizes in the Mie scattering regime (diameters between approximately 0.5-50~\textmu m given 55 Cancri e's host star's emission peaks near 0.5 \textmu m \citep{Bourrier2018b}). More detailed modeling of silicate cloud microphysics suggests growth to such sizes is plausible on sub-hourly to daily time scales if cloud condensation nuclei exist \citep{Powell2018}. Aerosols generated by magma vapor atmospheric chemistry can act as condensation nuclei for silicate cloud particles and are likely to be present on 55 Cancri e given the observed emission temperatures \citep{Schaefer2009,Lee2018}; if the magma ocean is not global, rocky surface-atmosphere interactions could supply additional aerosols to act as condensation nuclei \citep{Helling2019}. The time required for vapor parcels to grow cloud particles to radiatively relevant sizes is thus another potential driver of the lag $d$. While estimates for particle growth timescales appear consistent with the estimated range of $d$ used in this work, variable vapor supply due to surface temperature oscillations may drive changes in particle growth efficiency that are neglected here.

The heat capacity of the mixed layer of the surface magma ocean is poorly constrained. Moreover, it may significantly vary as surface heating and cooling change mixed layer thickness over an oscillation period. At high surface temperatures, the thin mixed layer and the resulting small heat capacity of the surface magma ocean could allow a rapid increase in surface temperature and thus a surge in the supply of silicate vapor. Therefore, the assumption of a constant mixed layer depth also likely results in smoother oscillations than in reality. Moreover, Kite et al. \cite{Kite2016} show that under some circumstances, the magma pool surface can be both time-variable and compositionally variegated (patchy).  A~dynamical magma ocean model coupled to surface forcings \cite[e.g.,][]{Lai2024a} will be needed to simulate the ocean mixed layer with better accuracy.

Additionally, our model neglects the longwave opacity of background gases in the atmosphere, despite recent evidence of its presence from absorption features in the 4.5~\textmu m band \cite{Hu2024}. This longwave opacity could potentially warm the magma. Atmospheric opacity at 4.5 \textmu m can also influence the 4.5~\textmu m brightness temperature. Whether this opacity increases or decreases the observed brightness temperature may depend on the vertical temperature gradient at the 4.5 \textmu m emission level.

Our model omits horizontal variations in surface temperature and cloud distribution. If different regions on the dayside undergo unsynchronized limit cycles, the averaged observable effect could be diminished. However, we consider such asynchrony unlikely, as the 0.2--2 km s$^{-1}$ winds \cite{Hammond2017,Nguyen2020} would likely distribute clouds and their cooling effects efficiently, thereby synchronizing the limit cycles across different regions.

%
\clearpage
\section*{Acknowledgements}
The project was made possible by the summer school \textit{Rossbypalooza: Clouds and convection in diverse climates}. \textit{Rossbypalooza} is funded by the University Corporation for Atmospheric Research (UCAR), National Science Fundation (NSF), and the Department of the Geophysical Sciences and Graduate Council at the University of Chicago. We thank Renyu Hu and an anonymous reviewer for helpful comments which improved the manuscript.

\section*{Author contributions}
Y.L., K.L., E.K., and B.F. designed research; K.L. and Y.L. performed research; K.L. and Y.L. analyzed data; Y.L. and K.L. designed graphics; Y.L. wrote the paper; K.L., E.K., and B.F. improved the paper.

\section*{Data, Materials, and Software Availability}

Data for the analysis in this study is uploaded to Zenodo: \url{https://doi.org/10.5281/zenodo.15122200}. Model and data visualization code are available at \url{https://github.com/kaitlyn-loftus/lava-limits}.

\section*{Declaration of interest}
The authors declare no conflict of interest.
\clearpage

\clearpage
\bibliography{bib}
\clearpage

%
{\noindent\huge\sffamily\bfseries Methods\par}\vspace{5mm}
\subsection*{Model Formulation}

Key elements in the magma temperature-cloud feedback are shown in Fig. \ref{fig:schematic}. Under clear-sky conditions, the stellar radiation flux received by a zero-albedo surface in the substellar region, $S_0 = 3.4 \times 10^6$ W~$\mathrm{m^{-2}}$, could sustain a magma ocean. We characterize the mean surface temperature of the substellar magma ocean with $T_{\text{surf}}$. Above the magma ocean, the atmosphere consists of non-condensable gases, such as $\mathrm{CO_2}$, CO, or $\mathrm{N_2}$. The magma ocean supplies vapor (e.g., SiO) derived from silicates, which can be transported to higher altitudes by substellar convection, where it subsequently condenses to form silicate clouds. The rate of vapor supply increases with $T_{\text{surf}}$. We focus on the surface temperature of the substellar region, rather than the mean surface temperature of the dayside, when formulating the cloud production rate below. We do so as the hottest substellar region is likely the primary source of cloud-forming vapor, given the strong temperature dependence of saturated vapor pressure.

For simplicity, we define ``shortwave'' radiation as the radiation from the star and ``longwave'' radiation as the radiation from the planet surface and cloud. We~assume that the cloud scatters shortwave radiation but does not absorb it. We denote the cloud optical depth in the shortwave as $\tau_{\text{SW}}$. For longwave radiation, we assume that the cloud absorbs and emits it but does not scatter it. We denote the cloud optical depth in the longwave as $\tau_{\text{LW}}$. Neglecting potential changes in cloud particle size, we assume that $\tau_{\text{SW}}$ and $\tau_{\text{LW}}$ are proportional to each other:
\begin{equation}\label{eq:taulinearity}
    \tau_{\text{LW}} = \beta \tau_{\text{SW}}
\end{equation}
where $\beta$ is a fixed parameter. $\tau_{\text{SW}}$ attenuates the incoming stellar radiation and cools the surface, while the longwave opacity induces a greenhouse effect that warms the surface and reduces the outgoing thermal radiation. The only exception to this simplification is that when diagnosing the planet's brightness temperature in the visible band, we assume that cloud attenuates surface-emitted visible light by scattering it but not absorbing it.

We formulate the evolution of shortwave cloud optical depth $\tau_{\text{SW}}$ by the following arguments. For the source, we assume that the conversion rate from cloud-forming vapor to cloud particles scales with the upward vapor flux across the cloud base. Considering a time delay (or “lag”) $d$ for vapor to ascend from the surface to the cloud base, the upward vapor flux at the cloud base at time $t$ scales with the near-surface upward vapor flux at $t-d$. This flux can be decomposed into the near-surface number density of vapor molecules and ascending velocity. According to the ideal gas law, the near-surface number density of vapor molecules scales with the $T_{\text{surf}}$-dependent partial pressure of the vapor divided by surface temperature $T_{\text{surf}}$. The ascending velocity of the vapor is assumed to scale with the atmospheric scale height divided by the vertical transport timescale, with the former being proportional to the mean atmospheric temperature $\Bar{T}$ and the latter proportional to the lag $d$. To summarize, the vapor flux across the cloud base is assumed to scale with
\begin{equation}\label{eq:vaporflux}
    \frac{N_{\text{A}} \Bar{T} p_{\text{ref}}}{g M_{\text{air}} T_{\text{surf}} d} \exp(-\frac{T_{\text{ref}}}{T_{\text{surf}}|_{t-d}}).
\end{equation}
where $N_{\text{A}}$ is Avogadro’s number, $g$ the gravitational acceleration, $M_{\text{air}}$ the mean molar mass of the background atmosphere, and $T_{\text{ref}}$ a reference temperature with a corresponding saturated vapor pressure $p_{\text{ref}}$. We denote the surface temperature at $t-d$ as $T_{\text{surf}}|_{t-d}$. A further simplification is to assume that $\Bar{T}/T_{\text{surf}}$ is approximately constant, given the large variability in the exponential term.

The exponential term is based on the Clausius-Clapeyron equation for phase transition. We obtain $p_{\text{ref}}$ and $T_{\text{ref}}$ by fitting $p_\text{ref}\exp(T_{\text{ref}}/T_{\text{surf}}|_{t-d})$ to the equilibrium vapor pressure at the magma ocean surface. For this fitting we use pressure values calculated by Kite et al. \cite{Kite2016}, which used the MAGMA chemical equilibrium code \cite{Fegley1987,Schaefer2004,Schaefer2009}. The reference pressure $p_{\text{ref}}$ and temperature $T_{\text{ref}}$ may vary with vapor and magma ocean composition. In this study, we use the values for SiO vapor above a magma ocean with a bulk silicate Earth composition: $p_{\text{ref}} = 8.1 \times 10^{13}$ Pa and $T_{\text{ref}} = 7.1 \times 10^4$ K. We choose SiO as it is among the most abundant cloud-forming vapors in the atmosphere above a magma ocean with a bulk silicate Earth composition \cite{Schaefer2009,Kite2016,Mahapatra2017}. Sensitivity tests show that our results are not qualitatively sensitive to the choice of cloud-forming vapor (Fig. \ref{fig:s4}).

Assuming a fixed fraction of vapor condenses as it ascends across the cloud base, the formation rate of condensates integrated over a cloud column, in units of molecules $\mathrm{m}^{-2}$~$\mathrm{s^{-1}}$, can be approximated as
\begin{equation}
    \frac{k_1 N_{\text{A}} p_{\text{ref}}}{g M_{\text{air}} d} \exp(-\frac{T_{\text{ref}}}{T_{\text{surf}}|_{t-d}})
\end{equation}
where $k_1$ is a dimensionless scale factor that accounts for uncertainties in this scaling, including those in relative humidity, pressure at the cloud base, rate of mixing during vapor transport, condensation fraction, etc.

To convert the formation rate of condensates into the rate of increase in cloud optical depth, we need to consider the microphysics involved. Assuming the molar mass of the vapor is $M_{\text{v}}$, we can derive the mass of condensates formed in a cloud column per unit time. With the density of the condensate $\rho_{\text{c}}$ and the mean radius of cloud particles $r_{\text{c}}$, the mean mass of a spherical cloud particle is $\frac{4}{3}\pi\rho_{\text{c}} r_{\text{c}}^3$, from which we can derive the production rate of cloud particles in a cloud column.

In the shortwave, the extinction cross-section of a single cloud particle is approximately $2\pi r_c^2$ \cite[][Chapter 5.4]{Pierrehumbert2010}. Therefore, the rate of increase in the extinction coefficient integrated over a cloud column, or the rate of increase in cloud optical depth, is
\begin{equation}
    \frac{3 k_1 M_{\text{v}} p_{\text{ref}}}{2 g M_{\text{air}} \rho_{\text{c}} r_{\text{c}} d} \exp(-\frac{T_{\text{ref}}}{T_{\text{surf}}|_{t-d}}).
\end{equation}

We assume the loss term of cloud optical depth scales with cloud optical depth itself, resulting in a ``residence time'' for cloud particles. We conservatively estimate this residence time assuming cloud loss occurs via particle sedimentation. The residence time is assumed to scale with the atmospheric scale height $R^\ast T/(M_{\text{air}} g)$ divided by the Stokes falling velocity $2\rho_{\text{c}} g r_{\text{c}}^2/(9\eta)$, where $\eta$ is the dynamic viscosity of the background atmosphere. For simplification, we use a constant characteristic atmospheric temperature $T_0$ in place of $T$ in the expression for atmospheric scale height. Therefore, the loss rate of cloud optical depth in the shortwave can be expressed as
\begin{equation}
    \frac{2 k_2 g^2 M_{\text{air}} \rho_{\text{c}} r_{\text{c}}^2}{9R^\ast T_0 \eta} \tau_{\text{SW}}
\end{equation}
where $k_2$ is a dimensionless coefficient accounting for uncertainties in the scaling.

To summarize, Eq. \ref{eqn:dtaudt} describes the time rate of change of shortwave cloud opacity $\tau_{\text{SW}}$:
\begin{equation}\label{eqn:dtaudt}
    \frac{\mathrm{d}\tau_{\text{SW}}}{\mathrm{d}t}=\frac{3k_1 M_{\text{v}} p_{\text{ref}}}{2g M_{\text{air}} \rho_{\text{c}} r_{\text{c}} d} \exp(-\frac{T_{\text{ref}}}{T_{\text{surf}}|_{t-d}}) -\frac{2 k_2 g^2 M_{\text{air}} \rho_{\text{c}} r_{\text{c}}^2}{9R^\ast T_0 \eta} \tau_{\text{SW}}.
\end{equation}
Thus, the variability in cloudiness is powered by fluctuations in the surface temperature of the substellar magma ocean. Cloud dissipation only responds to cloud formation, relaxing the system towards clear-sky conditions.

To model the evolution of $T_{\text{surf}}$, we consider radiative heating and cooling as well as the effect of latent heat during magma solidification and melting. Given the strong thermal emission at the high temperature, we neglect the effects of sensible heat and latent heat during magma vaporization. In the atmosphere, we only consider the longwave opacity of clouds and neglect the longwave opacity of background gases. Surface radiative heating is from the absorbed stellar shortwave radiation and the downward cloud longwave radiation. With a near-zero surface albedo, as supported by the observed near-zero secondary eclipse depths across visible to near-IR bands, the absorbed shortwave radiation is:
\begin{equation}
    \frac{S_0}{1+\alpha\tau_{\text{SW}}}
\end{equation}
where $S_0 = 3.4\times 10^6$ W m$^{-2}$ is the incoming stellar flux at the top of the atmosphere and $1/(1 + \alpha \tau_{\text{SW}})$ accounts for multiple scattering of the cloud. Sensitivity tests indicate our results are not qualitatively sensitive to the exact $\alpha$ value (Fig. \ref{fig:s4}), so, for simplicity, we choose a single intermediate value $\alpha=0.5$. No absorption by cloud is assumed in the shortwave, so the cloud albedo in the shortwave $A$, which measures the reflected fraction of the total incoming stellar radiation, is
\begin{equation}\label{eq:albedo}
    A = 1 - \frac{1}{1+\alpha\tau_{\text{SW}}}.
\end{equation}
This functional form approximates the derivation of planetary albedo in ref. \cite{Pierrehumbert2010} Chapter 5.6 for conservative scattering and no surface reflection.

We assume that the cloud radiates like a blackbody in the longwave, emitting both upward towards space and downward towards the surface. Given the potential vertical extent of the cloud, we model the cloud's upward emission at a temperature $T_{\text{cloud}\uparrow}$ with emissivity $\varepsilon_{\text{cloud}}$, and its downward emission at a different temperature $T_{\text{cloud}\downarrow}$ with the same emissivity $\varepsilon_{\text{cloud}}$. Modeling the vertical structure of the cloud is beyond the scope of this study, so for simplicity we introduce
\begin{equation}
    \Delta T_{\text{cloud}} = T_{\text{cloud}\downarrow} - T_{\text{cloud}\uparrow}
\end{equation}
as a model parameter to represent the cloud's vertical extent. A larger $\Delta T_{\text{cloud}}$ corresponds to a larger vertical extent.

Following Beer's law, we represent cloud emissivity as
\begin{equation}\label{eq:eps}
    \varepsilon_{\text{cloud}} = 1 - \exp(-\tau_{\text{LW}}) = 1 - \exp(-\beta\tau_{\text{SW}}).
\end{equation}
Assuming zero surface albedo in the longwave, the downward longwave radiative flux emitted by the cloud and received by the surface is then $\varepsilon_{\text{cloud}}\sigma T_{\text{cloud}\downarrow}^4$, where $\sigma$ is the Stefan-Boltzmann constant.

The surface cools by emitting longwave radiation as a blackbody at $T_{\text{surf}}$, with an emission flux given by $\sigma T_{\text{surf}}^4$. We assume that the cloud absorbs a portion of this flux with an absorptivity $\varepsilon_{\text{cloud}}$. Consequently, the energy balance equation for the cloud can be written as:
\begin{equation}\label{eq:cloud_radeq}
    \varepsilon_{\text{cloud}} \sigma T_{\text{surf}}^4 = \varepsilon_{\text{cloud}} \sigma T_{\text{cloud}\uparrow}^4 + \varepsilon_{\text{cloud}} \sigma T_{\text{cloud}\downarrow}^4.
\end{equation}
This equation indicates that the cloud's emission temperatures, $T_{\text{cloud}\uparrow}$ and $T_{\text{cloud}\downarrow}$, are determined by the variable surface temperature $T_{\text{surf}}$ and the static model parameter $\Delta T_{\text{cloud}}$.

We assume only a well-mixed surface layer of the magma ocean is thermally influenced by the oscillations. Thus, the evolution of surface temperature is determined by
\begin{equation}\label{eqn:dTdt}
    \frac{\mathrm{d}T_{\text{surf}}}{\mathrm{d}t} = \frac{S_0/(1 + \alpha \tau_{\text{SW}}) + \varepsilon_{\text{cloud}}\sigma T_{\text{cloud}\downarrow}^4 - \sigma T_{\text{surf}}^4}{c_p \rho_{\text{m}} H_{\text{mix}}}
\end{equation}
where $c_p$ is the specific heat capacity of the magma, $\rho_{\text{m}}$ the magma density, and $H_{\text{mix}}$ the thickness of the surface mixed layer. The product $c_p \rho_{\text{m}} H_{\text{mix}}$ is the heat capacity of the surface magma ocean. To account for latent heat release during magma solidification, we do not allow $T_{\text{surf}}$ to fall below the solidus temperature $T_{\text{solidus}}$ = 1400 K \cite{Katz2003} while surface net energy loss occurs. Accordingly, to account for latent heat absorption during magma melting, we hold $T_{\text{surf}}$ at $T_{\text{solidus}}$ while surface net energy gain occurs until the previously released latent heat is fully absorbed before increasing $T_{\text{surf}}$ above $T_{\text{solidus}}$. For the parameters considered here, we find the magma ocean's surface layer never completely solidifies.

Combining Eqs. \ref{eqn:dTdt} and \ref{eqn:dtaudt}, we have formulated a closed set of equations that allows us to predict the behavior of the $T_{\text{surf}}$-$\tau_{\text{SW}}$ system. We then nondimensionalize time $t$ against the lag time $d$, so any potential oscillation period can be expressed as multiples of $d$. Denoting the dimensionless time as $t^\prime=t/d$, Eqs. \ref{eqn:dtaudt} and \ref{eqn:dTdt} can then be rewritten as
\begin{equation}\label{eqn:dtaudtprime}
    \frac{\mathrm{d}\tau_{\text{SW}}}{\mathrm{d}t^\prime} = \frac{3k_1 M_{\text{v}} p_{\text{ref}}}{2g M_{\text{air}} \rho_{\text{c}} r_{\text{c}}} \exp(-\frac{T_{\text{ref}}}{T_{\text{surf}}|_{t^\prime-1}}) -\frac{2 k_2 g^2 M_{\text{air}} \rho_{\text{c}} r_{\text{c}}^2 d}{9R^\ast T_0 \eta} \tau_{\text{SW}}
\end{equation}
and
\begin{equation}\label{eqn:dTdtprime}
    \frac{\mathrm{d}T_{\text{surf}}}{\mathrm{d}t^\prime}= \frac{S_0/(1 + \alpha \tau_{\text{SW}}) + \left[1 - \exp(-\beta\tau_{\text{SW}})\right]\sigma T_{\text{cloud}\downarrow}^4 - \sigma T_{\text{surf}}^4}{c_p \rho_{\text{m}} H_{\text{mix}}/d},
\end{equation}
respectively.

We define three bulk model parameters:
\begin{equation}
    \Pi_1=\frac{3k_1 M_{\text{v}}}{2g M_{\text{air}} \rho_{\text{c}} r_{\text{c}}},
\end{equation}
\begin{equation}
    \Pi_2=\frac{2k_2 g^2 M_{\text{air}} \rho_{\text{c}} r_{\text{c}}^2 d}{9R^\ast T_0 \eta},
\end{equation}
and
\begin{equation}
    \Pi_3 = \frac{c_p \rho_{\text{m}} H_{\text{mix}}}{d}.
\end{equation}
$\Pi_1$ may be interpreted as a shortwave scatterer (or longwave absorber) productivity, $\Pi_2$ as the loss efficiency of shortwave scatterer (or longwave absorber), and $\Pi_3$ as a thermal inertia. Eqs. \ref{eqn:dtaudtprime} and \ref{eqn:dTdtprime} can then be rewritten as
\begin{equation}\label{eq:dtaudtprime_nond}
    \frac{\mathrm{d}\tau_{\text{SW}}}{\mathrm{d}t^\prime} = \Pi_1 p_{\text{ref}} \exp(-\frac{T_{\text{ref}}}{T_{\text{surf}}|_{t^\prime-1}})-\Pi_2 \tau_{\text{SW}}
\end{equation}
and
\begin{equation}\label{eq:dTdtprime_nond}
    \frac{\mathrm{d}T_{\text{surf}}}{\mathrm{d}t^\prime} = \frac{S_0/(1 + \alpha \tau_{\text{SW}}) + \varepsilon_{\text{cloud}}\sigma T_{\text{cloud}\downarrow}^4 - \sigma T_{\text{surf}}^4}{\Pi_3},
\end{equation}
respectively.

We then analyze the solutions to the equations above (with the additional constraint that $T_{\text{surf}}$ cannot fall below $T_{\text{solidus}}$) for various values of the bulk model parameters $\Pi_1$, $\Pi_2$, and $\Pi_3$, as well as $\beta$ and $\Delta T_{\text{cloud}}$. Through numerical simulations, we identify regions in the parameter space where $T_{\text{surf}}$ and $\tau_{\text{SW}}$ exhibit oscillations that can induce the observed variability in secondary eclipse depths. The parameter space is explored within the bounds listed in Table \ref{tab:varied_param}. The limits for the bulk parameters $\Pi_1$, $\Pi_2$, and $\Pi_3$ are derived from the estimated bounds of individual model parameters (such as $\rho_{\text{c}}$, $r_{\text{c}}$, $c_p$, and $H_{\text{mix}}$, see Supporting Information). We set the upper bound for $\Delta T_{\text{cloud}}$ based on the difference between the atmosphere’s emission temperature and the stratospheric skin temperature, assuming the former is equal to the highest observed 4.5 \textmu m brightness temperature from ref. \cite{Patel2024}. Because the same condensate mass contributes to shortwave and longwave radiative effects, $\beta$ is controlled by the ratio of longwave to shortwave extinction efficiencies, a function of cloud particle size, wavelength-dependent condensate optical properties, and cloud temperature \citep[e.g.,][]{Hansen1974,Pierrehumbert2010}. For a given particle size, extinction efficiency largely tends to increase or stay the same as wavelength decreases. However, if the particle radius is approximately the longwave wavelength, longwave extinction efficiency could be a factor of a few larger than shortwave. Therefore, we explore $10^{-3}<\beta<4$ to represent varying degrees of cloud longwave emissivity, along with $\beta=0$, which corresponds to no cloud longwave emissivity. With $\beta=0$, $\Delta T_{\text{cloud}}$ has no effect in the model, reducing the five-dimensional parameter space to three-dimensions (spanned solely by $\Pi_1$, $\Pi_2$, and $\Pi_3$).

\subsection*{Numerical Method for Equation Solving}

We solve Eqs. \ref{eq:dtaudtprime_nond} and \ref{eq:dTdtprime_nond} for $T_{\text{surf}}$ and $\tau_{\text{SW}}$ with a stiffly-accurate fourth-order explicit singly diagonal implicit Runge-Kutta (ESDIRK) method \cite{Kennedy2003} using the Julia differential equation solving ecosystem DifferentialEquations.jl \cite{Rackauckas2017,Widmann2022}. We use adaptive time stepping with relative tolerance of $10^{-8}$ and absolution tolerance of $10^{-10}$. Tests with lower tolerances suggest our results are numerically converged (Table \ref{tab:S2}). Between $t^\prime=0$ and $t^\prime=1$, we fix $T_{\text{surf}}|_{(t^\prime-1)}$ at the initial temperature value but let $\tau_{\text{SW}}$ and $T_{\text{surf}}$ vary according to the two equations.

We find that for the parameter space of Table \ref{tab:varied_param} (except when $\beta>1$), all solutions either reach a steady state at their fixed point ($T_{\text{surf}}^\ast$, $\tau_{\text{SW}}^\ast$), which depends on the model parameters, or oscillate on a stable limit cycle on the phase diagram (see an example in Fig. \ref{fig:limitcycleex}). For a given parameter set, we solve for the fixed point ($T_{\text{surf}}^\ast$, $\tau_{\text{SW}}^\ast$) numerically by setting the left-hand sides of Eqs. \ref{eq:dtaudtprime_nond} and \ref{eq:dTdtprime_nond} equal to zero. When the solution leads to a limit cycle, this fixed point is unstable with the limit cycle encircling it.

During integration, we check if a solution is in steady state, in a limit cycle, or neither, by considering integration steps within the previous 100 units of the nondimensionalized time $t^\prime$, which is longer than any limit cycle periods we detect. If the maximum and minimum surface temperatures within this duration are equal to $T_{\text{surf}}^\ast$ within $\pm$1 K, we consider the solution in steady state. If not, we search for all times $T_{\text{surf}}$ crosses $T_{\text{surf}}^\ast$. From these $T_{\text{surf}}^\ast$ crossings, we calculate periods of oscillations about $T_{\text{surf}}^\ast$. We test if these periods significantly vary, or specifically, if the standard deviation of all periods calculated over the previous 100 units of $t^\prime$ is less than 1\% of the mean period $P$. If we find a consistent period, we consider the solution to be in a limit cycle if $\tau(t^\prime) \approx \tau(t^\prime + P)$ within $10^{-2}$ relative error and $T_{\text{surf}}(t^\prime) \approx T_{\text{surf}}(t^\prime + P)$ within $10^{-3}$ relative error  over the previous 100 units of $t^\prime$. In all other scenarios, we consider the solution to still be evolving. 

When $\beta>1$, we find a small fraction of solutions do not reach steady state or enter into a limit cycle (as defined above) after an extended integration period (3000 $d$). These solutions show erratic variability, in some cases possibly consistent with chaotic behavior (Supporting Information, Fig. \ref{fig:erratic_example}-\ref{fig:betaevo}). We conservatively exclude solutions exhibiting this behavior from model-observation comparison.  

\subsection*{Brightness Temperature Calculation and Model-Observation Comparisons}

We investigate what portion of the plausible parameter space of Table \ref{tab:varied_param} can lead to limit-cycle oscillations that are consistent with the observed variability in the 4.5 \textmu m brightness temperature \cite{Demory2016a,Tamburo2018,Patel2024}. To efficiently sample this five-dimensional parameter space, we use Latin hypercube sampling \cite{McKay2000} of log $\Pi_1$, log $\Pi_2$, log $\Pi_3$, $\Delta T_{\text{cloud}}$, and log $\beta$ within the bounds presented in Table \ref{tab:varied_param}. Latin hypercube sampling divides each parameter range into equally spaced bins, randomly samples one point from each bin, and then randomly combines samples of different parameters to generate parameter sets. For each parameter set, we solve the equations for $T_{\text{surf}}$ and $\tau_{\text{SW}}$ until their solution reaches a final state, either a steady state or a limit cycle.

At 4.5~\textmu m (in the longwave), reflection by cloud is negligible, so we calculate the brightness temperature for the substellar region $T_{\text{b,4.5}}$ following
\begin{equation}\label{eqn:TbrightLW}
    B(T_{\text{b,4.5}},4.5) = B(T_{\text{surf}},4.5)\mathrm{e}^{-\tau_{\text{LW}}} + B(T_{\text{cloud}\uparrow},4.5) \varepsilon_{\text{cloud}},
\end{equation}
where $B(T,\lambda)$ is the Planck function for a blackbody with temperature $T$ at wavelength $\lambda$ (\textmu m).

We apply Eq. \ref{eqn:TbrightLW} to final-state time series of $T_{\text{surf}}$, $\tau_{\text{LW}}$, and $T_{\text{cloud}\uparrow}$, and calculate the lowest and the highest $T_{\text{b,4.5}}$ for each parameter combination. We then assess if a parameter combination can produce a solution consistent with observations by comparing the modeled highest and lowest $T_{\text{b,4.5}}$ with the observed values \cite{Demory2016a,Tamburo2018,Patel2024}. Considering the reported uncertainties in the observed brightness temperatures, we classify the parameter space into 1$\sigma$, 2$\sigma$, and 3$\sigma$ tiers, corresponding to agreement with both the deepest and shallowest secondary eclipse depth observations within one, two, and three times the measurement errors, respectively. The highest disk-mean brightness temperature of the dayside at 4.5 \textmu m is no lower than 2448 K, 2080 K, and 1712 K within one, two, and three times the measurement errors, respectively, as reported by the fifth data point in Table 4 of ref. \cite{Demory2016a}. These lower bounds on the highest disk-mean brightness temperature of the dayside can also serve as lower bounds for the highest brightness temperature of the substellar region, as the latter is probably higher. The lowest disk-mean brightness temperature of the dayside at 4.5 \textmu m is no higher than 1040 K, 1207 K, and 1374 K within one, two, and three times the measurement errors, respectively, as indicated by the first data point in Table 1 of ref. \cite{Patel2024}. However, these upper bounds can not be directly applied to the probably higher lowest brightness temperature of the substellar region. The maximum difference between the substellar brightness temperature and the disk-mean brightness temperature of the dayside might occur on a 1:1 tidally locked planet without an atmosphere to redistribute heat. On such a planet, assuming a fully absorbing Lambertian surface, the substellar surface temperature is $\sim$10\% higher than the disk-mean surface temperature of the dayside. Using this as a reference, we adjust the upper bounds on the lowest brightness temperature of the substellar region by a factor of 1.1, resulting in values of 1144 K, 1328 K, and 1511 K for 1$\sigma$, 2$\sigma$, and $3\sigma$ observational consistency, respectively.

Due to the challenge of distinguishing between the contributions of reflected starlight and the planet’s thermal emission in the observed brightness temperature in the visible and near-IR bands, we do not assess the consistency of model outputs with observations based on the brightness temperatures in these bands \cite{MeierValdes2022,Demory2023,Patel2024}. However, we compute the modeled brightness temperature in the visible band to illustrate that it can oscillate out of phase with the modeled brightness temperature in the mid-IR (e.g., at 4.5 \textmu m wavelength). The brightness temperature at 0.5~\textmu m, $T_{\text{b,0.5}}$, is calculated using
\begin{equation}\label{eqn:TbrightSW}
    B(T_{\text{b,0.5}},0.5) = AB(T_{\text{star}},0.5)\frac{R_{\text{star}}^2}{a^2} + (1 - A)B(T_{\text{surf}},0.5),
\end{equation}
where $T_{\text{star}}=5172$ K is the effective temperature of the host star \cite{Bourrier2018b}, $R_{\text{star}}=6.56\times10^8$ m is the radius of the host star \cite{VonBraun2011}, $a=2.33\times10^9$ m is the semi-major axis of the planet's orbit \cite{Dawson2010}, and $A$ is the shortwave albedo in Eq. \ref{eq:albedo}. The first term represents the reflected starlight, while the second term accounts for thermal emission from the planetary surface that penetrates through the cloud. At visible wavelengths, clouds attenuate surface-emitted radiation by scattering it, which is why the cloud's shortwave albedo is included here.
\clearpage

\setcounter{figure}{0}
\setcounter{table}{0}
\setcounter{equation}{0}
\renewcommand{\thetable}{S\arabic{table}}
\renewcommand{\thefigure}{S\arabic{figure}}
\renewcommand{\theequation}{S\arabic{equation}}
{\noindent\huge\sffamily\bfseries Supplementary Materials\par}\vspace{5mm}
\section*{Role of lag $\bm{d}$ in variability}
In this section, we (numerically) demonstrate that non-steady state solutions (i.e., variability) require a non-zero time lag $d$ for vapor at the surface to become condensed cloud particles in the atmosphere. In order to test the behavior of our model with no lag, we need to reformulate the governing equations to remove division by $d$. We return to our model's governing equations before non-dimensionalization by $d$: main text Eqs. (\ref{eqn:dtaudt}) and (\ref{eqn:dTdt}). We define three bulk parameters from these equations, analogous to the three bulk parameters defined from the equations with non-dimensionalized time:  
\begin{equation}
    \Pi_1^\prime=\frac{3k_1 M_{\text{v}}}{2g M_{\text{air}} \rho_{\text{c}} r_{\text{c}}\tau_\text{ascent}},
\end{equation}
\begin{equation}
    \Pi_2^\prime=\frac{2k_2 g^2 M_{\text{air}} \rho_{\text{c}} r_{\text{c}}^2}{9R^\ast T_0 \eta},
\end{equation}
and
\begin{equation}
    \Pi_3^\prime = c_p \rho_{\text{m}} H_{\text{mix}}.
\end{equation}
Note, $\tau_\text{ascent}$ in the expression for $\Pi_1^\prime$ is a timescale for vertical transport. Previously we assumed this timescale to be $d$. For the purpose of this analysis, we emphasis the role of $d$ in coupling the magnitude of vapor supply to an earlier surface temperature rather than its secondary role in $\Pi_1$ as the supply timescale. (A more complex model could separate these timescales.) Here to prevent an unphysical infinite vapor supply, we assume $\tau_\text{ascent}= 1$ hour from the lower bound on $d$. 
We estimate the possible ranges of the revised bulk parameters as $1.75 \times 10^{-7}$ -- $7.09 \times 10^{-2}$ Pa$^{-1}$ s$^{-1}$ for $\Pi_1^\prime$, $1.27 \times 10^{-11}$ -- $5.94 \times 10^{-3}$ s$^{-1}$ for $\Pi_2^\prime$, and $2.38 \times 10^{6}$ -- $6.61 \times 10^{9}$ kg s$^{-2}$ K$^{-1}$ for $\Pi_3^\prime$.

Setting the lag to zero in the Clausius-Clapeyron expression of $\dv*{\tau_\text{SW}}{t}$ (i.e., $T_{\text{surf}}|_{t-d}=T_{\text{surf}}|_t$) and substituting the new bulk parameters, main text Eqs. (\ref{eqn:dtaudt}) and (\ref{eqn:dTdt}) can be rewritten as
\begin{equation}\label{eq:dtaudt_nodelay}
    \frac{\mathrm{d}\tau_{\text{SW}}}{\mathrm{d}t} = \Pi_1^\prime p_{\text{ref}} \exp(-\frac{T_{\text{ref}}}{T_{\text{surf}}})-\Pi_2^\prime \tau_{\text{SW}}
\end{equation}
and
\begin{equation}\label{eq:dTdt_nodelay}
    \frac{\mathrm{d}T_{\text{surf}}}{\mathrm{d}t} = \frac{S_0/(1 + \alpha \tau_{\text{SW}}) + \varepsilon_{\text{cloud}}\sigma T_{\text{cloud}\downarrow}^4 - \sigma T_{\text{surf}}^4}{\Pi_3^\prime},
\end{equation}
respectively.
We solve Eqs. (\ref{eq:dtaudt_nodelay})-(\ref{eq:dTdt_nodelay}) numerically and determine their end state dynamical behavior (steady state or limit cycle) following the procedure described in the main text (``Numerical Method for Equation Solving''). The only difference is that we use an explicit fifth order Runge-Kutta method to integrate \citep{Tsitouras2011}. 

We run $10^5$ parameter combinations generated via Latin hypercube sampling of log $\Pi_1^\prime$, log $\Pi_2^\prime$, log $\Pi_3^\prime$, $\Delta T_\text{cloud}$, and $\beta$. We find all parameter combinations reach steady state and show no variability. Therefore, we conclude that non-zero lag $d$ is required for our model to generate variability.

We provide the physical intuition for why the lag $d$ cannot be zero or too small for oscillations to occur. The lag $d$ effectively holds back the restoring force in the system—the formation of clouds that cool the surface—while the driving force—the rapid vapor release from a hot surface—remains active. The driving force pushes the system away from equilibrium, and if $d$ is too small, the restoring force is released too soon, pulling the system back before it has moved far from equilibrium. This limits the oscillation amplitude and can even prevent oscillations altogether.

\section*{Solutions not converging to steady state or detectable limit cycle}

\begin{figure}
\centering
\includegraphics[width=\textwidth]{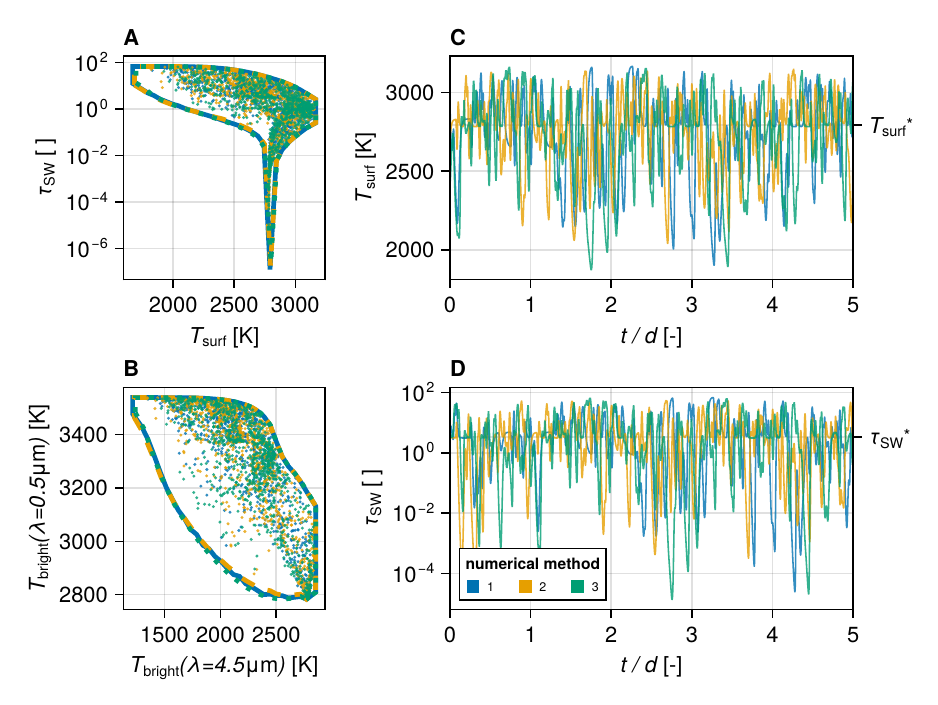}
\caption{An example of a model simulation not converging to steady state or a detectable limit cycle. Colors indicate three different numerical methods (as labeled in panel $D$): (1) our default numerical setup (blue), (2) our default numerical setup except the numerical solver's relative tolerance is set 10\% lower (yellow), and (3) our default numerical setup except we use a different method for integration \citep[green;][]{Kvaerno2004}. ($A$) Phase space of shortwave optical depth $\tau_\text{SW}$ vs surface temperature $T_\text{surf}$. Scatter points show 1,000 random points from the last 2,000 $d$ of simulation. The thick line encircling the scatter points shows the maximum extent of the phase space. (In addition to color, different numerical methods also have different line styles to see their overlap.) The three methods approximately share the same attractor. ($B$) Like Panel $A$ except phase space for the 0.5~\textmu m brightness temperature $T_{\text{b,0.5}}$ vs 4.5~\textmu m brightness temperature $T_{\text{b,4.5}}$. ($C$) $T_{\text{surf}}$ vs non-dimensionalized time $t/d$. ($D$) $\tau_\text{SW}$ vs $t/d$. Model parameters: $\Pi_1=1.75$ $\mathrm{Pa}^{-1}$, $\Pi_2=377$, $\Pi_3=28.8$ $\mathrm{kg}$ $\mathrm{s}^{-3}$ $\mathrm{K}^{-1}$, $\Delta T_{\text{cloud}}=323$ $\mathrm{K}$, and $\beta=3.31$.}\label{fig:erratic_example}
\end{figure}

When $\beta>1$, we encounter a small fraction of parameter combinations (about 2\%) in the parameter space spanned by Figure \ref{fig:parambounds} that do not converge to our numerical definitions of steady state or a limit cycle even after running the model 10$\times$ longer than the majority of cases take to converge (3000$d$ vs 300$d$). Figure \ref{fig:erratic_example} shows an example of one of these non-converging solutions encountered while constructing Figure \ref{fig:parambounds}. The three colors show three numerical approaches: (1) our default numerical setup, (2) our default numerical setup except the numerical solver's relative tolerance is set 10\% lower, and (3) our default numerical setup except we use a different method for integration \citep{Kvaerno2004}. All three numerical approaches show consistent erratic variability, which suggests this solution property is likely not due to our numerical implementation. 

It is computationally prohibitive to integrate large numbers of these erratic solutions much past $3000d$ due to high memory requirements from over 10 million numerical integrator steps. (We tested different numerical solvers but did not identify one that could produce significantly fewer steps.) Therefore, we conservatively exclude solutions not converging to steady state or a detectable limit cycle from consideration of observational consistency in Figure \ref{fig:parambounds}. Inspection of multi-dimensional parameter dependencies and numerical convergence tests indicate that short period detected limit cycles ($P \lesssim 0.8 d$) are also associated with this variability. We therefore conservatively exclude limit cycles with $P < 0.8 d$ when $\beta > 1$ as well. These exclusions have very minor implications for our results as they exclude less than 0.2\% of solutions from inclusion in Figure \ref{fig:parambounds}.

Even though our coupled system of equations is two dimensional, because our model includes a delay differential equation, it can exhibit dynamical behaviors of higher dimensional systems, including possibly chaos \citep{Wernecke2019}. Of future interest is further examining whether the magma temperature-cloud feedback could produce chaotic behavior when $\beta = \tau_\text{LW} / \tau_\text{SW} > 1$. Such behavior could also potentially explain the seemingly uncorrelated near-IR and mid-IR brightness temperatures observed by \cite{Patel2024} (c.f. Fig. \ref{fig:erratic_example}B). 

\begin{figure}
\centering
\includegraphics[width=\textwidth]{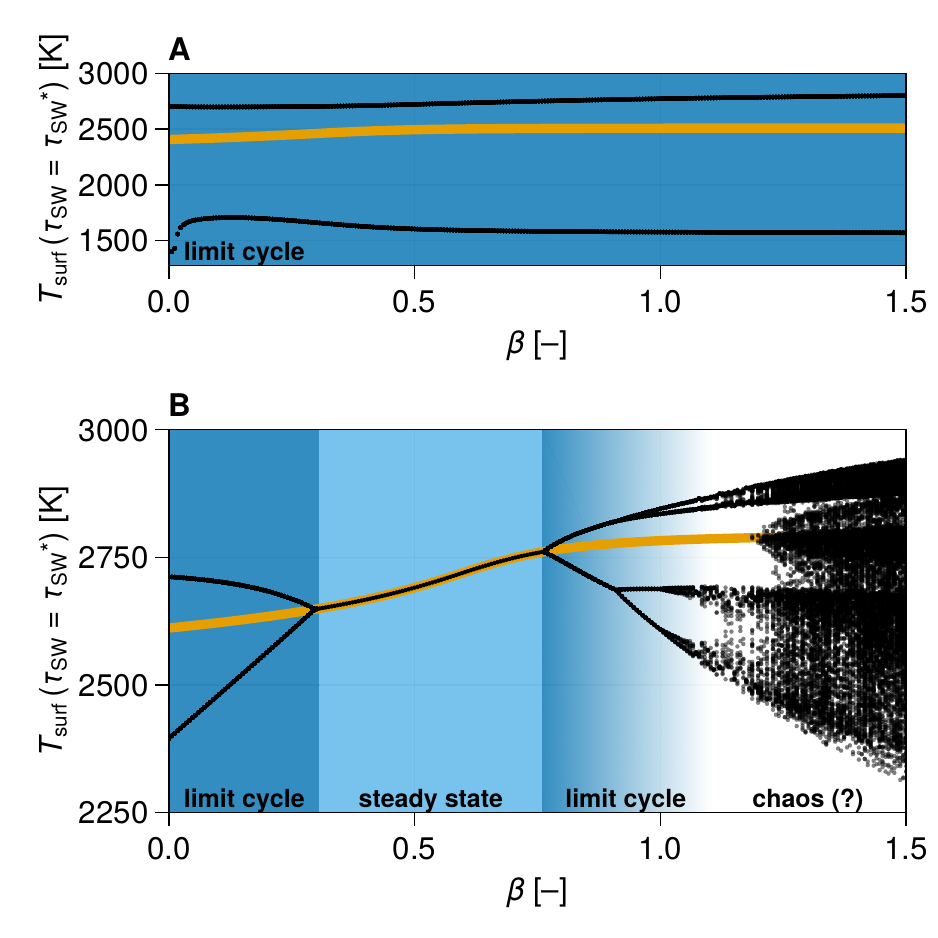}
\caption{Example bifurcation diagrams for variable $\beta$ (ratio of longwave optical depth to shortwave optical depth) and all other model parameters fixed. Black scatter points show values of surface temperature $T_\text{surf}$ when shortwave optical depth $\tau_\text{SW}$ is equal to its fixed point $\tau_\text{SW}^\ast$ for the final 50$d$ of simulation vs $\beta$. The solid yellow line shows the surface temperature fixed point $T_\text{surf}^\ast$ vs $\beta$. As $\beta$ varies, the system can switch dynamical regimes (indicated by different background color shading as labeled). Panel $A$ shows a typical limited dependence on $\beta$ across most of the parameter space explored. Panel B shows an example of a rare case where solutions begin to behave erratically as $\beta$ increases. After a series of period doubling bifurcations, chaotic behavior appears to occur for $\beta \gtrsim 1.05$. Panel $A$ model parameters: $\Pi_1=9.47$ $\mathrm{Pa}^{-1}$, $\Pi_2=74.2$, $\Pi_3=115$ $\mathrm{kg}$ $\mathrm{s}^{-3}$ $\mathrm{K}^{-1}$, and $\Delta T_{\text{cloud}}=177$ $\mathrm{K}$. Panel $B$ model parameters: $\Pi_1=1.75$ $\mathrm{Pa}^{-1}$, $\Pi_2=377$, $\Pi_3=28.8$ $\mathrm{kg}$ $\mathrm{s}^{-3}$ $\mathrm{K}^{-1}$, and $\Delta T_{\text{cloud}}=323$ $\mathrm{K}$. Note, these diagrams only show behavior when $\beta$ varies but all other models parameters are fixed; therefore, they are not reflective of Figure \ref{fig:parambounds} (which varies in 5 dimensions rather than 1).}\label{fig:bifurcation_diagram}
\end{figure}

In Figure \ref{fig:bifurcation_diagram}, we examine the dynamical regime of our model as a function of $\beta$ for two sets of other model parameters. 
For each value of $\beta$, we plot values of $T_\text{surf}$ when $\tau_\text{SW}$ is equal to its fixed point $\tau_\text{SW}^\ast$ over the last 50$d$ of simulation. (The simulation terminates due to reaching convergence to an end state (steady state or limit cycle) or reaching 3000$d$.) In steady state, $T_\text{surf}$($\tau_\text{SW}^\ast$) has one unique value, the fixed point $T_\text{surf}^\ast$. For a ``simple'' limit cycle, $T_\text{surf}$($\tau_\text{SW}^\ast$) has two unique values straddling $T_\text{surf}^\ast$. More than two unique values of $T_\text{surf}$($\tau_\text{SW}^\ast$) for a given parameter set suggests a more complex limit cycle or chaos.

\begin{figure}
\centering
\includegraphics[width=\textwidth]{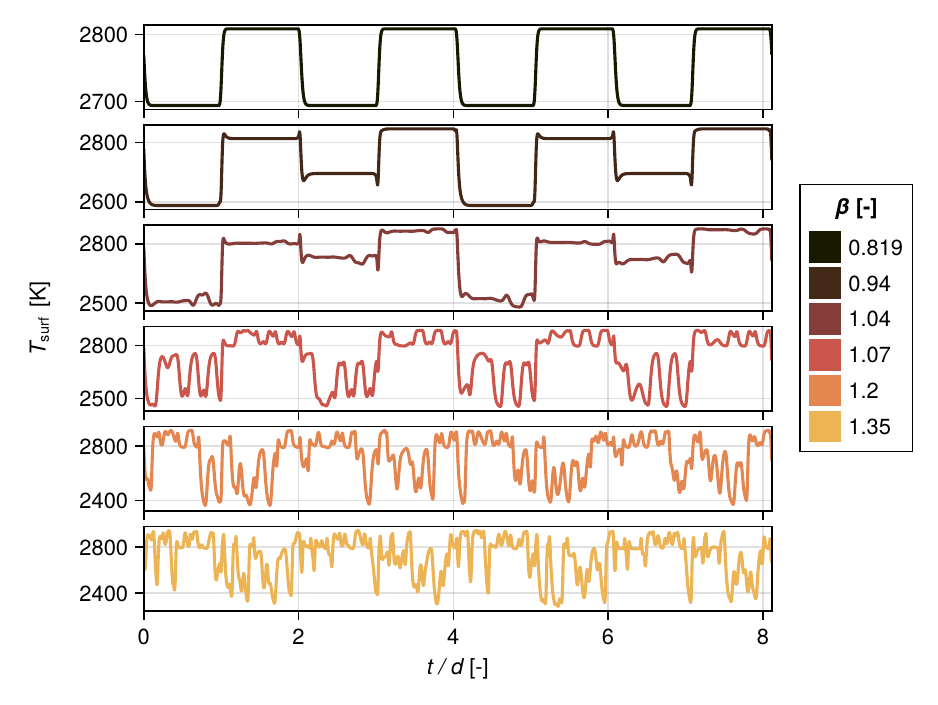}
\caption{Surface temperature $T_\text{surf}$ vs non-dimensionalized time $t/d$ for six different values of $\beta$ (ratio of longwave optical depth to shortwave optical depth) and fixed other parameters. Values of $\beta$ are labeled by color and increase downward. The time series duration shown is four periods of the top row limit cycle ($\beta=0.819$). Other model parameters (same as Figure \ref{fig:bifurcation_diagram}: $\Pi_1=1.75$ $\mathrm{Pa}^{-1}$, $\Pi_2=377$, $\Pi_3=28.8$ $\mathrm{kg}$ $\mathrm{s}^{-3}$ $\mathrm{K}^{-1}$, and $\Delta T_{\text{cloud}}=323$ $\mathrm{K}$.}\label{fig:betaevo}
\end{figure}

The parameter set visualized in Figure \ref{fig:bifurcation_diagram}A shows a typical behavior of our model as $\beta$ varies: the dynamical regime does not change. The parameter set visualized in Figure \ref{fig:bifurcation_diagram}B shows a rare behavior of our model: dynamical regime sensitivity to $\beta$ with erratic variability emerging for $\beta>1$.  
Figure \ref{fig:betaevo} shows time series of $T_\text{surf}$ for six of the $\beta$ values in Figure \ref{fig:bifurcation_diagram}B.    
\clearpage




\begin{figure}
\centering
\includegraphics[width=0.75\textwidth]{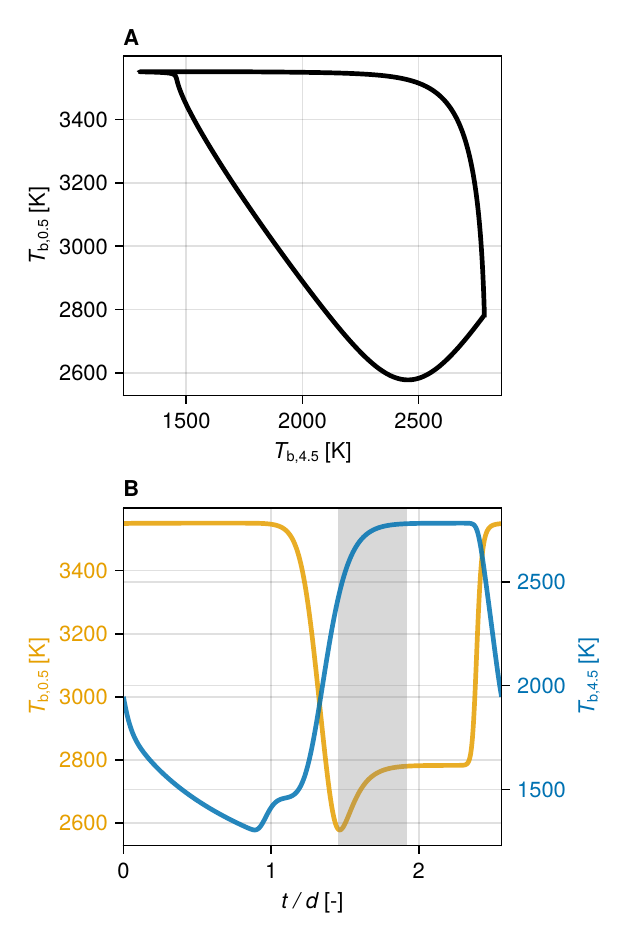}
\caption{An example limit cycle with an out-of-phase oscillation qualitatively similar to the JWST/NIRCam observations of \citep{Patel2024} (see Fig. \ref{fig:JWST}). ($A$) Phase diagram of the 0.5~\textmu m brightness temperature $T_{\text{b,0.5}}$ and the 4.5~\textmu m brightness temperature $T_{\text{b,4.5}}$. ($B$) Time series (in non-dimesionalized time $t / d$) of $T_{\text{b,0.5}}$ (yellow, left vertical axis) and $T_{\text{b,4.5}}$ (blue, right vertical axis) over one oscillation period. The gray shading highlights the time of the out-of-phase oscillation. Model parameters: $\Pi_1=93.3$ $\mathrm{Pa}^{-1}$, $\Pi_2=18.8$, $\Pi_3=439$ $\mathrm{kg}$ $\mathrm{s}^{-3}$ $\mathrm{K}^{-1}$, $\Delta T_{\text{cloud}}=28.5$ $\mathrm{K}$, and $\beta=2.36 \times 10^{-3}$.}\label{fig:s1}
\end{figure}
\clearpage

\begin{figure}
\centering
\includegraphics[width=\textwidth]{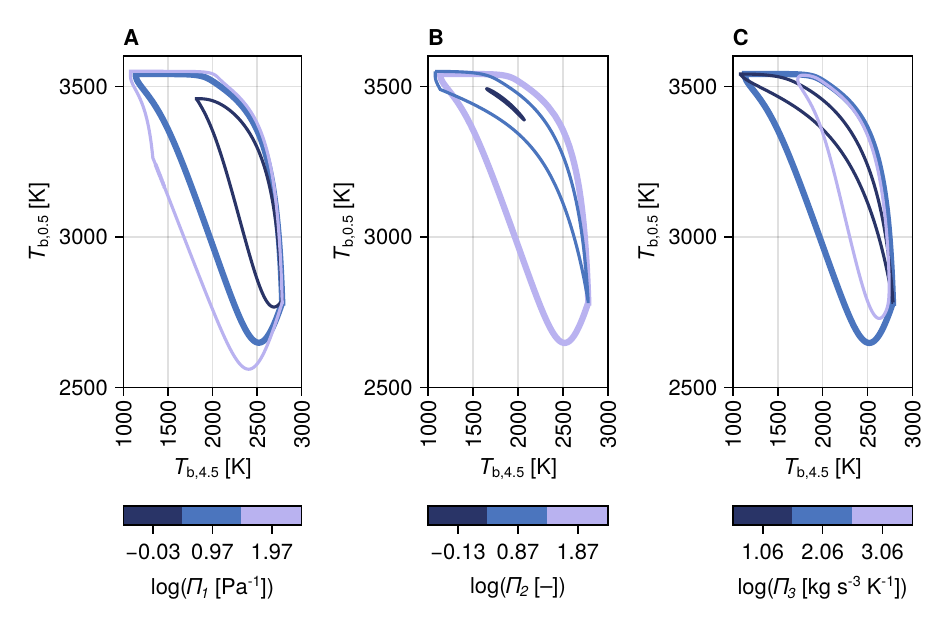}
\caption{Figure \ref{fig:limitcycleex}B (phase diagram of the 4.5~\textmu m brightness temperature $T_{\text{b,4.5}}$ and the 0.5~\textmu m brightness temperature $T_{\text{b,0.5}}$) with perturbed bulk parameters: ($A$) shortwave scatterer (or longwave absorber) productivity $\Pi_1$, ($B$) loss efficiency of shortwave scatterer (or longwave absorber) $\Pi_2$, and ($C$) thermal inertia $\Pi_3$. In each panel, three curves are shown only differing by their $\Pi$ value (as labeled in colorbar) with the thickest line representing the default parameter value used in Figure \ref{fig:limitcycleex}. Model parameters (unless modified as indicated by colorbar) follow Figure \ref{fig:limitcycleex}: $\Pi_1=9.47$ Pa$^{-1}$, $\Pi_2=74.2$, $\Pi_3=115$ kg s$^{-3}$ K$^{-1}$, $\Delta T_{\text{cloud}}=177$ K, and $\beta=0.143$.}\label{fig:S2}
\end{figure}
\clearpage

\begin{figure}
\centering
\includegraphics[width=\textwidth]{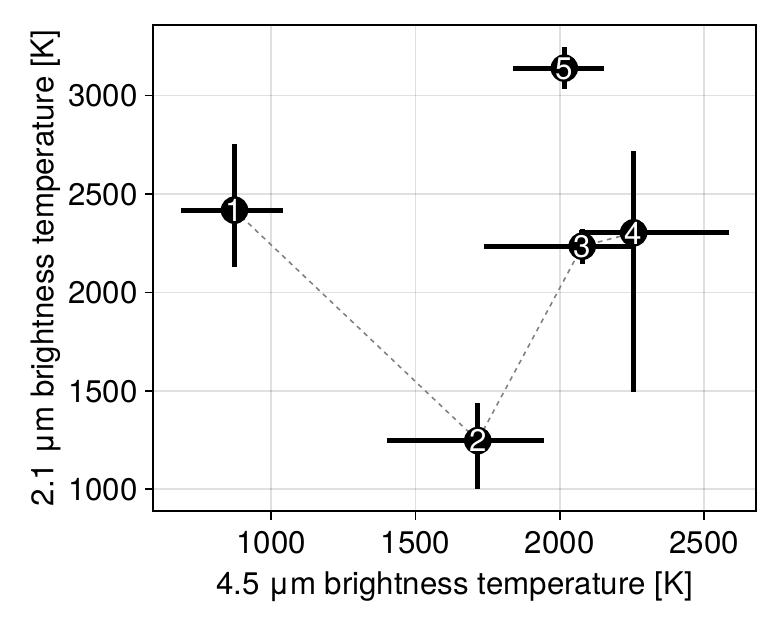}
\caption{JWST/NIRCam 2.1 \textmu m brightness temperature vs 4.5 \textmu m brightness temperature from Patel et al. \cite{Patel2024} Table 1. Error bars correspond to 1 standard deviation in retrieved brightness temperature. Numeric labels on scatter points correspond to “visit” eclipse number. Dashed lines connect the four eclipse measurements taken within a week.}\label{fig:JWST}
\end{figure}
\clearpage

\begin{figure}
\centering
\includegraphics[width=\textwidth]{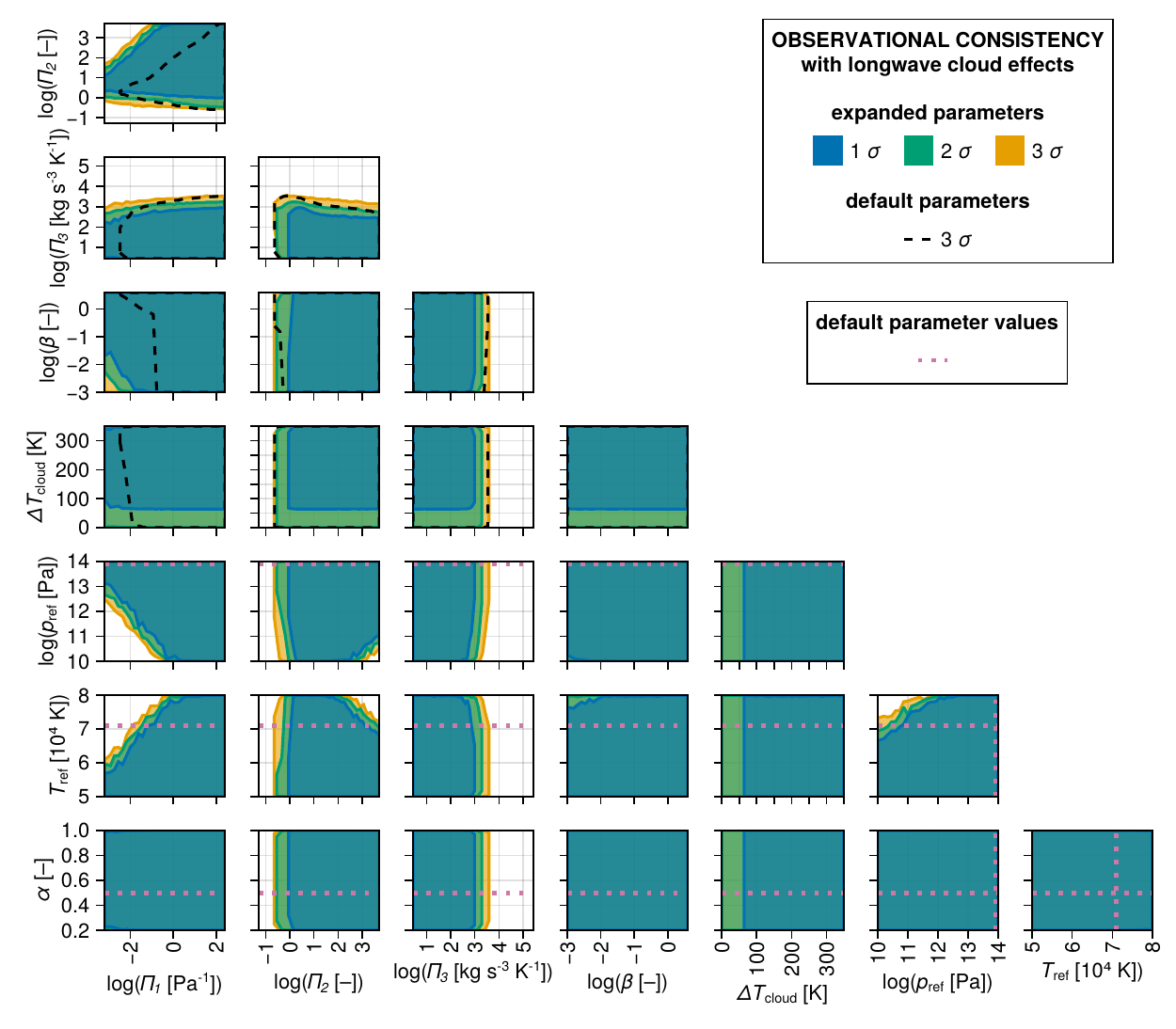}
\caption{Similar to Figure \ref{fig:parambounds} with additional parameter sensitivity tests for three parameters fixed in the main text: reference pressure $p_\text{ref}$, reference temperature $T_\text{ref}$, and albedo parameter $\alpha$. Parameter ranges capable of simultaneously reproducing the upper and lower bounds of Spitzer and JWST 4.5 \textmu m brightness temperature observations within one, two, and three times the measurement uncertainty ($\sigma$). The shading and solid contours show the model with longwave cloud effects with additional $p_\text{ref}$, $T_\text{ref}$, and $\alpha$ parameter perturbations. Different colors indicate different levels of consistency. For comparison, the dashed black contours in the first four rows show the model (with longwave cloud effects) as presented in Figure \ref{fig:parambounds} at $3\sigma$ model-observation consistency. The dotted pink lines in the last three rows show the default parameter values of $p_\text{ref}$, $T_\text{ref}$, and $\alpha$ in Figure \ref{fig:parambounds}.}\label{fig:s4}
\end{figure}
\clearpage

\clearpage

\clearpage

\clearpage 

 \begin{table}\centering
 \caption{Estimated bounds for individual model parameters that constitute $\bm{\Pi_1}$, $\bm{\Pi_2}$, and $\bm{\Pi_3}$}\label{tab:S1}

 \begin{tabular}{c p{5cm} l p{5.3cm}}
 Parameter & Physical meaning & Value or bounds & Note \\
 \midrule
 $R^\ast$ & universal gas constant & 8.3144 $\mathrm{J}$ $\mathrm{mol^{-1}}$ $\mathrm{K^{-1}}$ &  \\
 $g$ & gravitational acceleration & 22 $\mathrm{m}$ $\mathrm{s^{-2}}$ & From ref. \cite{Bourrier2018b} \\
 $M_{\text{v}}$ & molar mass of the cloud-forming vapor & 44 $\mathrm{g}$ $\mathrm{mol^{-1}}$ & For $\mathrm{SiO}$ \\
 $M_{\text{air}}$ & mean molar mass of the background atmosphere & 28--44 $\mathrm{g}$ $\mathrm{mol^{-1}}$ & Considering CO, $\mathrm{N_2}$, and $\mathrm{CO_2}$, supported by ref. \cite{Hu2024} \\
 $\rho_{\text{c}}$ & density of cloud particles & 2100--3600 $\mathrm{kg}$ $\mathrm{m^{-3}}$ & Lower bound: for $\mathrm{SiO}$, upper bound: for $\mathrm{Mg_2SiO_4}$ \\
 $r_{\text{c}}$ & mean radius of cloud particles & 0.2--30 $\mathrm{\mu m}$ & From ref. \cite{Powell2018} \\
 $T_0$ & representative atmospheric temperature & 675--3000 $\mathrm{K}$ & Lower bound: stratospheric skin temperature based on the lowest observed nightside brightness temperature, upper bound: the highest surface temperature \\
 $\eta$ & dynamic viscosity of the background atmosphere & $4.6\times10^{-5}$--$8.0\times10^{-5}$ $\mathrm{kg}$ $\mathrm{m^{-1}}$ $\mathrm{s^{-1}}$ & Lower bound: for $\mathrm{N_2}$ or CO at 1500 K, upper bound: for $\mathrm{CO_2}$ at 3000 K, from refs. \cite{Weast1986,Crane1988} \\
 $c_p$ & specific heat capacity of magma & 1090--2360 $\mathrm{J}$ $\mathrm{kg^{-1}}$ $\mathrm{K^{-1}}$ & Lower bound: for $\mathrm{FeO}$ melt, upper bound: for $\mathrm{MgO}$ melt, from ref. \cite{Lange1992} \\
 $\rho_{\text{m}}$ & density of magma & 2180--2800 $\mathrm{kg}$ $\mathrm{m^{-3}}$ & Lower bound: for rhyolitic magma, upper bound: for basaltic magma \\
 $H_{\text{mix}}$ & thickness of the mixed layer of the surface magma ocean & 1--1000 $\mathrm{m}$ & Upper bound referenced from refs. \cite{Lai2024a,Lai2024b} \\
 $d$ & time delay between high surface temperature and accelerated cloud formation & 1 hour to 10 days & Lower bound: surface-to-cloud vertical transport timescale, upper bound: horizontal transport timescale to encircle the planet \\
 $k_1$ & scale factor accounting for uncertainties in the scaling for the increase rate of cloud optical depth & 0.001--1 & Lower bound: limited supersaturation, particle nucleation barriers, subsaturated near-surface atmosphere, mixing with ambient air \\
 $k_2$ & scale factor accounting for uncertainties in the scaling for the residence time of cloud & 0.1--100 & Lower bound: non-spherical particles \cite{Ohno2020}, cloud depth uncertainty,
upper bound: particle growth, cloud depth uncertainty \cite{Rossow1978} \\
 \bottomrule
 \end{tabular}
 \end{table}
\clearpage

\begin{table}\centering
 \caption{Numerical convergence test}\label{tab:S2}
 \begin{tabular}{llr}
  summary statistic & quantity & magnitude of relative error \\
 \midrule
  maximum & min($T_\text{surf}$) & $2.48 \times 10^{-3}$\\
  mean & min($T_\text{surf}$) & $1.50 \times 10^{-5}$\\
  maximum & max($T_\text{surf}$) & $1.27 \times 10^{-3}$\\
  mean & max($T_\text{surf}$) & $7.51 \times 10^{-6}$\\
  maximum & min($T_{\text{b,4.5}}$) & $2.56 \times 10^{-3}$\\
  mean & min($T_{\text{b,4.5}}$) & $1.81 \times 10^{-5}$\\
  maximum & max($T_{\text{b,4.5}}$) & $1.06 \times 10^{-3}$\\
  mean & max($T_{\text{b,4.5}}$) & $8.39 \times 10^{-6}$\\
  \bottomrule
 \end{tabular}
 
 \addtabletext{To perform a numerical convergence test, we model $10^{5}$ parameter combinations generated via Latin hypercube sampling of log $\Pi_1$, log~$\Pi_2$, log $\Pi_3$, $\Delta T_\text{cloud}$, and $\beta$ twice: once with our default numerical setup and once with 10\% of the default relative and absolute tolerance. The magnitude of relative error (|RE|) for quantity $x$ is calculated from a model simulation run with default numerics (subscript ``default'') and a model simulation (with the same parameters) run with 10\% of the default tolerances (subscript ``10\% tol.'') via
 
 \begin{equation}
 \text{|RE|} = \left|\frac{x_\text{default} - x_\text{10\% tol.}}{x_\text{10\% tol.}}\right|.
 \end{equation}
 
 The summary statistic (mean or maximum) is calculated from all runs converging to steady state or detectable limit cycle in our default numerical setup (as required for inclusion in Figure \ref{fig:parambounds}). (We exclude the erratic variability runs as we do not necessarily believe they have converged.) The quantity is calculated over the last $100 d$ of simulation (i.e., the same procedure for generating Figure \ref{fig:parambounds}).}  
\end{table}
\clearpage








\clearpage

\end{document}